\documentclass[
  preprintnumbers,amsmath,amssymb]{revtex4}
  \usepackage{graphicx}

\begin{document}
\title{Nonergodic Brownian oscillator}
\author{Alex V. Plyukhin}
\email{aplyukhin@anselm.edu}
 \affiliation{ Department of Mathematics,
Saint Anselm College, Manchester, New Hampshire 03102, USA 
}

\date{\today}

\begin{abstract}
We consider an open (Brownian) classical harmonic oscillator
in contact with a non-Markovian thermal bath and described by the generalized Langevin equation.
When the bath's spectrum has a finite upper cutoff frequency,
the oscillator may have  ergodic   and nonergodic configurations.
In ergodic configurations (when  exist, they correspond to lower oscillator frequencies)
the oscillator demonstrates conventional relaxation to thermal equilibrium with the bath.
In nonergodic configurations (which correspond to higher oscillator frequencies)
the oscillator in general does not thermalize, but relaxes to periodically correlated (cyclostationary)
states whose statistics vary periodically in time.
For a specific dissipation kernel in the Langevin equation, we evaluate explicitly relevant relaxation functions, 
which 
describe the evolution of mean values and time correlations.
When the oscillator frequency is switched 
from a lower value to higher one, 
the oscillator
may show parametric ergodic to nonergodic transitions with  equilibrium initial  and  cyclostationary final states.
These transitions are shown to resemble phase transitions of the second kind.

\end{abstract}


\maketitle

\section{Introduction: Two mechanisms of non-thermalization}

Since thermodynamics is a very general theory, the cases when it does not apply in the ordinary way are intriguing.
A thermodynamic description assumes that a system, large or small, in contact with a macroscopic thermal bath at
temperature $T$ will thermalize over the course of time, reaching a state of thermodynamic equilibrium characterized
by the same temperature $T$. 
While this
scenario is the most  common, 
there are certain types of open systems for which
thermalization does not occur. 
For a classical particle in contact with the thermal bath
we are aware of two mechanisms of relaxation which does not end up with thermal equilibrium.
The first mechanism is due to the zero integral friction~\cite{Costa,Bao,Lapas,Morgado,Siegle},
the second is related  to the formation of a localized vibrational
mode~\cite{Montroll,Teramoto,Kashiwamura,Rubin,Flach,loc_mode,MM,Dhar,Onofrio,Wei,Ishikawa,Plyukhin1,Plyukhin2}.
While this paper concerns exclusively the latter, let us start with a brief  outline of the former.

Consider a free Brownian particle of mass $m$ described by the generalized Langevin equation~\cite{Zwanzig}
\begin{eqnarray}
\dot v(t) =-\int_0^t K(t-\tau) \,v(\tau)\, d\tau +\frac{1}{m}\,\xi(t),
\label{le0}
\end{eqnarray}
where $v$ is the particle's velocity, $K(t)$ is the dissipation kernel, and $\xi(t)$ is
the zero-centered stationary random force, connected to $K(t)$ by the
conventional fluctuation-dissipation relation. 
Assuming the random force does not correlate with $v(0)$,  one finds from Eq. (\ref{le0}) that
the normalized correlation function $R(t)=\langle v(t) v(0)\rangle/\langle v^2(0)\rangle$
satisfies the homogeneous (and deterministic) equation
\begin{eqnarray}
\dot R(t) =-\int_0^t K(t-\tau) \,R(\tau)\, d\tau
\label{C_eq}
\end{eqnarray}
with the initial condition $R(0)=1$. In the Laplace domain
the solution is
\begin{eqnarray}
\tilde R(s)=\int_0^\infty e^{-st} R(t)\,dt=\frac{1}{s+\tilde K(s)}.
\end{eqnarray}
Thermalization implies that the system eventually forgets initial conditions,
so that $R(t)$ vanishes at long times. On the contrary,  the  condition of 
non-thermalization  implies that $R(t)$ does not vanish at long times, which leads to the asymptotic 
condition on the dissipation kernel in the Laplace domain
\begin{eqnarray}
\lim_{t\to\infty} R(t)=\lim_{s\to 0} s\,\tilde R(s)=\lim_{s\to 0}\frac{s}{s+\tilde K(s)}\ne 0,
\label{cond_lim}
\end{eqnarray}
provided the limit exists.
This condition is satisfied when the Laplace transform of the kernel has the asymptotic form
\begin{eqnarray}
\tilde K(s)\sim s^\delta, \quad \delta\ge 1, \quad s\to 0.
\label{cond0}
\end{eqnarray}
In that case the integral of the kernel vanishes
\begin{eqnarray}
\gamma=\int_0^\infty K(t)\, dt=\tilde K(s=0)=0.
\label{cond00}
\end{eqnarray}
This may be called the condition of zero integral friction, because
$\gamma=\int_0^\infty K(t) dt$ is just the friction (dissipation) 
coefficient  in  the expression for the damping force $-\gamma\, p$
in the Markovian limit of the Langevin equation (\ref{le0}), $\dot v=-\gamma\, v+m^{-1}\xi$.

The lack of thermalization in the case $\gamma=0$ is remarkable but appears to be a rather exotic phenomenon.
One can show that under condition (\ref{cond0})
Brownian motion shows not only the lack of thermalization but also another anomalous phenomenon,
namely super-diffusion (which actually takes place for a
broader condition $\delta>0$)~\cite{Costa,Bao,Lapas,Morgado,Siegle}.

At first glance it may appear that the condition  of non-thermalization
(\ref{cond0}) is both sufficient and necessary.
Actually it is not necessary because the above discussion  assumes, when the
relation $\lim_{t\to\infty} R(t)=\lim_{s\to 0} s \tilde R(s)$ is exploited,
that the correlation function $R(t)$ possesses a well-defined  long time limit.
That is not necessarily  the case in general:
It is easy to construct 
a dissipation  kernel
$K(t)$ with reasonable properties (such as $K(t)\to 0$ at long times) for which
Eq. (\ref{C_eq}) for $R(t)$ has an oscillating solution and $\lim_{t\to\infty} R(t)$ does not exists~\cite{Plyukhin0}.
Can such  mathematical possibility be realized in any physical system?

One such system is well-known: it is the harmonic lattice with a light impurity atom.
Such atom can be viewed as a 
lattice defect which is known to generate a localized vibrational mode whose
frequency (we shall denote it  $\omega_*$) lies outside the spectrum 
of the unperturbed lattice  
~\cite{Montroll,Teramoto,Kashiwamura,Rubin}. 
The localized mode involves the impurity atom and a few neighboring atoms of the lattice,
the participation of other atoms is small and decreases exponentially with the distance from the impurity.
The formation of the localized mode
can be attributed to destructive wave interference and is analogous to localization of the electron wave
function near the  impurity in the otherwise ideal crystal, see \cite{loc_mode} for a pedagogical discussion. For a review of  
localized modes in anharmonic lattices (called breathers) see~\cite{Flach}.
The localized mode does not exchange energy with  the bath; as a result  the light impurity atom (the system)
does not reach thermal equilibrium with the lattice (the bath).

In earlier studies,  localized vibrational modes and their unusual relaxation properties
were  studied by direct  solving  the equations of motion of the lattice.
Later the topic was addressed using the generalized Langevin equation (\ref{le0}), which often offers a more compact,
though less detailed, consideration~\cite{MM,Onofrio,Wei,Ishikawa,Dhar,Plyukhin1,Plyukhin2}.  
Within that method, 
the condition of non-thermalization due to localized modes
does not imply conditions (\ref{cond0}) and (\ref{cond00}),  yet it puts strong 
restrictions on the properties of the bath and the system-bath coupling.  
As mentioned above, the frequency of the localized mode lies outside the spectrum of the bath,
which necessarily implies that the latter must have a finite upper cutoff frequency $\omega_0$.  
That condition is satisfied neither for Markovian models with $K(t)\sim \delta(t)$, nor
for models with monotonically (e.g., exponentially) decaying  $K(t)$.

Non-thermalization due to localized  modes was demonstrated not only for a light
isotope in the harmonic lattice, but for a number of other models~\cite{MM,Onofrio,Wei,Ishikawa,Dhar}.  
While the lattice-like structure of 
the bath is probably the necessary ingredient (which guarantees that the  bath spectrum has a finite  upper bound $\omega_0$),
the system of interest may be of different nature. The earlier studies mostly concerned  lattice models with mass
and spring defects and their combinations.  
More recently,
non-thermalization of the Brownian  oscillator (both linear and nonlinear)
in the presence of localized  modes
was demonstrated by Dhar and Wagh~\cite{Dhar}.

The Brownian oscillator in contact with a Markovian bath and described by the standard Markovian Langevin
or Fokker-Planck equations is an exemplary  system, whose relaxation to thermal equilibrium can be analytically
described in full details. However, when the bath is not Markovian and has the frequency
spectrum with an upper cutoff $\omega_0$, the oscillator may thermalize or not thermalize depending on the values
of the oscillator frequency $\omega$ and parameters of the oscillator-bath coupling.
In the model studied in~Ref. \cite{Dhar}, the  oscillator thermalizes  for $\omega\le\omega_c$ and  does not thermalize
for $\omega>\omega_c$, where $\omega_c$ is a critical frequency of order of $\omega_0$. 
In configurations with $\omega>\omega_c$ 
the oscillator evolves at long times into 
a non-equilibrium and non-stationary state in which the mean values and correlation functions of dynamical variables 
oscillate with time  with the localized normal mode frequency $\omega_*$.

Stochastic processes with periodically varying statistics are called
periodically correlated,  or cyclostationary~\cite{Yaglom,Gardner1,Gardner2,Serpedin}. 
They are present
in a great variety of physical, biological, meteorological, and  technological processes,
involving an interplay of randomness and periodicity.
While both ingredients are obviously present in the Brownian oscillator,  the emergence of
cyclostationary states instead of stationary (equilibrium) states is rather unexpected from a thermodynamics point of view. 
Cyclostationary stochastic processes are not stationary and are therefore manifestly nonergodic:
their ensemble and time averages cannot be equal since the the former are time periodic and the latter are constant. 
If the oscillator, due to the formation of a localized mode, does not thermalize,
then it evolves in  a cyclostationary and hence  nonergodic state.

Since the type of relaxation may depend on the oscillator frequency $\omega$, we shall use the following nomenclature.
We will say that the oscillator with a given frequency is in an \textit{ergodic} configuration
if the oscillator relaxes to thermal equilibrium.  If the oscillator with 
a given frequency  does not thermalize (with the exception of equilibrium initial conditions)
but evolves in a cyclostationary state,  we shall say that the oscillator is in a \textit{nonergodic}
configuration. Based on the results of ~\cite{Dhar} one would expect ergodic (resp.~nonergodic)
configurations to correspond to lower (resp.~higher) oscillator frequencies.
We call a Brownian oscillator nonergodic if it has \textit{both} ergodic and nonergodic configurations,
or \textit{only} nonergodic ones. The oscillator with only ergodic configurations
(which will hardly appear in this text) may be referred as ergodic.

Note that, in general, the condition of thermalization is stronger than that of ergodicity
and non-thermalization does not necessarily imply nonergodicity. 
But for the nonergodic Brownian oscillator, this is indeed the case:
If the oscillator at given frequency fails to thermalize, then it will evolve
to a non-stationary (cyclostationary) and  therefore nonergodic state.

The purpose of this paper is to evaluate 
explicitly the relaxation and correlation functions describing 
a nonergodic Brownian oscillator.
Compared to the work by Dhar and Wagh~\cite{Dhar}, where  spectral properties of the bath are not specified
(except the part addressing a nonlinear oscillator), in this paper 
we shall focus on a case study of the oscillator described by the generalized Langevin equation with a  specific 
dissipation kernel $K(t)$. We believe such a study, though not generic,  would be of interest since it may share  many essential features
and technicalities with a variety of similar and extended models.
With explicit expressions for relaxation and correlation functions 
at hands, one can explore quantitatively a variety of nonergodic processes. 
As an application, we consider (in Sec. XIV) the optical trap like setting when the oscillator frequency
is instantaneously switched from a lower to higher value, bringing the oscillator from an ergodic to nonergodic configuration.
In that transition, the  nonergodic oscillator 
may get and, in contrast to its ergodic counterpart (whose energy eventually relaxes to $k_BT$),
store forever an arbitrary amount of energy.
That property may be of interest  for designing microscopic machines~\cite{BE,BE2}.

We have given the paper a mostly linear structure, with only two Appendices. For the reader's convenience,
Secs.~VIII through X, which  are fairly technical,   conclude with brief summaries that suffice for comprehending the following sections. 
The results of Secs.~III and IV are generic and do not depend on a specific form of the  dissipation
kernel $K(t)$; in the rest of the paper  the kernel $K(t)$ is adopted in the form (\ref{K}).


\section{Model}
We consider 
a classical Brownian particle of mass $m$ and coordinate $q$, trapped in the  harmonic potential
$V(q)=-m\omega^2\,q^2/2$, and 
in contact with a single thermal bath with temperature $T$.
The particle's
dynamics is governed by the generalized Langevin equation~\cite{Zwanzig}
\begin{eqnarray}
\ddot q(t)=-\omega^2\,q(t)-\int_0^t K(t-\tau)\,\dot q(\tau)\, d\tau+\frac{1}{m}\xi(t).
\label{gle}
\end{eqnarray}
The stationary fluctuating force (the noise) $\xi(t)$ is zero-centered and related
to the dissipation kernel $K(t)$
via the standard fluctuation-dissipation relation, 
\begin{eqnarray}
\langle \xi(t)\rangle=0, \qquad \langle \xi(t)\,\xi(\tau)\rangle=m\,k_B\,T\,K(|t-\tau|).
\label{fdr}
\end{eqnarray}
We shall assume that the kernel has a specific form,
\begin{eqnarray}
K(t)=\frac{\mu\,\omega_0^2}{4} [J_0(\omega_0 t)+J_2(\omega_0 t)]=\frac{\mu\,\omega_0}{2}\,\frac{J_1(\omega_0 t)}{t},
\label{K}
\end{eqnarray}
where $\mu$ and $\omega_0$ are arbitrary positive parameters,
$J_n(x)$ are Bessel functions of the first kind, and 
the second expression  is defined at $t=0$ by continuity.
The  kernel (\ref{K}) oscillates and decays at long times rather slowly as $t^{-3/2}$.
Otherwise, it has all properties one expects from the dissipation kernel of the generalized
Langevin equation: the function $K(t)$ given by (\ref{K}) is even,  has a maximum at $t=0$, and vanishes at long times. 

There are two reasons to pay a special attention to the kernel  (\ref{K}).
First, it is one of the simplest kernels for which the Langevin equation may have periodically correlated solutions. Indeed,  
the corresponding spectral density
\begin{eqnarray}
J(\omega)=\int_0^\infty K(t) \cos(\omega t)\,dt=
\frac{\mu\, \omega_0}{2}\, \sqrt{1-\left(\frac{\omega}{\omega_0}\right)^2}\, 
\theta(\omega_0-\omega),
\end{eqnarray}
where $\theta(\omega)$ is the step function, has the upper cutoff bound $\omega_0$.
As was mentioned in Introduction, this is expected to be the necessary condition of 
the localized mode formation and nonergodic configurations.
The model with the dissipation kernel decaying according to the power law,
$K(t)=K_0 t^{-\alpha}$, was considered earlier in Refs.~\cite{VD,DV}.
In that case the oscillator is ergodic and  thermalizes for any values of the oscillator frequency $\omega$.

The second special feature of the kernel (\ref{K}) is that it corresponds
to a specific and familiar physical model, namely Rubin's model, 
where the thermal bath
is the infinite  harmonic chain of atoms of mass $m_0$ and the system of interest is an isotope
atom of mass $m$~\cite{Rubin,Weiss,Zwanzig}. 
The Langevin equation (\ref{gle}) describes the original Rubin's model modified by the presence of the external
harmonic potential applied to the impurity atom.
The parameter $\omega_0$ has the meaning of the highest normal mode frequency of the infinite chain, 
$\omega_0=2\sqrt{k/m_0}$, where $k$ is the stiffness of the spring force
connecting atoms of the chain (and also the impurity atom).
There are two versions of Rubin's model. In the first version the isotope of mass $m$  
is attached to the end of the semi-infinite chain of atoms of mass $m_0$, see Fig. 3.1 in Ref.~\cite{Weiss}.
In that case the parameter $\mu$  has the meaning of the mass ratio $\mu=m_0/m$.  In the second version,
the isotope is embedded in the bulk of the infinite chain, i.e. attached to two semi-infinite chains.
For that version $\mu$ is the  doubled mass ratio, $\mu=2m_0/m$. The connection to Rubin's
model facilitates  computer simulations,  which may be helpful for
extended models (e.g., the oscillator is nonlinear, the oscillator frequency is subjected to a time variation, etc.)
when an analytical solution of the Langevin equation is not feasible. 

For the given model, it will be shown that for $\mu<2$ the oscillator has both ergodic  and nonergodic configurations.
Ergodic configurations correspond to lower frequencies 
\begin{eqnarray}
\omega\le \omega_c, \quad \omega_c=\sqrt{1-\mu/2}\,\omega_0,
\end{eqnarray}
while nonergodic  configurations correspond to higher frequencies $\omega>\omega_c$.  
On the other hand, for $\mu\ge 2$ the oscillator is nonergodic and does not thermalize for any frequency $\omega$. 
Our goal is to evaluate the relaxation and correlation functions (defined in the next two sections)
describing the oscillator's dynamics for both ergodic and nonergodic configurations.

\section{Solving Langevin equation}
The solution of  the generalized Langevin equation (\ref{gle}) 
for the harmonic oscillator 
with an arbitrary kernel $K(t)$
has been addressed
in several studies, see~\cite{Wang,VD,DV,Goychuk}. To make the paper self-contained we outline in this section the main points.

Solving Eq. (\ref{gle}) with initial conditions
$q(0)=q_i$ and $v(0)=v_i$ 
using the method of Laplace transform one finds for the coordinate and velocity of the particle the following expressions:
\begin{eqnarray}
q(t)&=&q_i\,S(t)+v_i\, G(t)+
\frac{1}{m}\,\{G*\xi\}(t),\nonumber \\
v(t)&=&-q_i\,\omega^2 G(t)+v_i\, R(t)+
\frac{1}{m}\,\{R*\xi\}(t).
\label{solutions}
\end{eqnarray}
Here
the asterisk denotes  the convolutions, e. g.
\begin{eqnarray}
\{G*\xi\}(t)=\int_0^t G(t-\tau)\, \xi(\tau)\, d\tau,
\label{conv}
\end{eqnarray}
the relaxation functions $G(t)$ is defined by its Laplace transform
\begin{eqnarray}
\tilde G(s)=\int_0^\infty e^{-st}\, G(t)\,dt=
\frac{1}{s^2+s\tilde K(s)+\omega^2},
\label{G}
\end{eqnarray}
and the other two relaxation functions $R(t)$ and $S(t)$ 
are derived from $G(t)$ as follows:
\begin{eqnarray}
R(t)=\frac{d}{dt} G(t),\qquad S(t)=1-\omega^2\int_0^t G(\tau)\, d\tau.
\label{RH}
\end{eqnarray}
As obvious from Eq. (\ref{solutions}), the initial values of the relaxation functions are  
\begin{eqnarray}
G(0)=0, \qquad R(0)=S(0)=1.
\label{GHR0}
\end{eqnarray}
Note that $G(t)$ has the dimension of time, while $R(t)$ and $S(t)$ are dimensionless.

In the Laplace domain the relaxation functions are connected as
\begin{eqnarray}
\tilde R(s)=s\,\tilde G(s), \quad \tilde S(s)=\frac{1}{s}\,[1-\omega^2 \tilde G(s)].
\label{relaxation_laplace}
\end{eqnarray}
Using  Eqs. (\ref{G}) and (\ref{relaxation_laplace})
one can directly verify the validity of relations
\begin{eqnarray}
s\, \tilde G(s)=\tilde S(s)-\tilde K(s)\,\tilde G(s),
\quad\quad
s\, \tilde R(s)-1=-\omega^2\, \tilde G(s)-\tilde K(s)\,\tilde R(s).
\label{laplace_relations}
\end{eqnarray}
In the time domain they give expressions for derivatives of $G$ and $R$:
\begin{eqnarray}
\dot G(t)=S(t)-\{K*G\}(t),
\quad\quad
\dot R(t)=-\omega^2 G(t)-\{K*R\}(t).
\label{GH_relations}
\end{eqnarray}
The derivative of $S$, according to Eq. (\ref{RH}), is 
\begin{eqnarray}
\dot S(t)=-\omega^2 G(t). 
\label{S_dot}
\end{eqnarray}

As follows from Eq. (\ref{solutions}),   the  relaxation to thermal equilibrium implies
the asymptotic vanishing of the relaxation functions 
\begin{eqnarray}
G(t), R(t), S(t)\to 0,\quad\mbox{as}\quad t\to\infty,
\label{cond_ergo}
\end{eqnarray}
which guarantees that the particle forgets the  initial  conditions  at long times. 
One can show that  the second moments of $q$ and $v$  under conditions (\ref{cond_ergo}) relax to the equilibrium values.
Indeed, by squaring solutions (\ref{solutions}), averaging over the initial parameters $q_i, v_i$, and assuming that 
\begin{eqnarray}
\langle q_i\rangle=\langle v_i\rangle=\langle q_i\, v_i\rangle=0
\label{IC1}
\end{eqnarray}
one gets the following expressions for the second moments
\begin{eqnarray}
\langle q^2(t)\rangle&=&\langle q_i^2\rangle\,S^2(t)+\langle v_i^2\rangle\, G^2(t)+
\frac{1}{m^2}\,\langle \{G*\xi\}^2(t)\rangle,\nonumber\\
\langle v^2(t)\rangle&=&\langle q_i^2\rangle\, \omega^4\, G^2(t)+\langle v_i^2\rangle\, R^2(t)+
\frac{1}{m^2}\,\langle \{R*\xi\}^2(t)\rangle,\nonumber\\
\langle q(t)\, v(t)\rangle&=&-\langle q_i^2\rangle\, \omega^2 G(t) S(t)+\langle v_i^2\rangle\,G(t)\, R(t)+\frac{1}{m^2}\,\langle \{G*\xi\}\,\{ R*\xi\}\rangle.
\label{moments1}
\end{eqnarray}
Here the average squares of the convolutions can be readily evaluated using the fluctuation-dissipation relation (\ref{fdr})
and also relations (\ref{GH_relations}) and (\ref{S_dot}),  
\begin{eqnarray}
\langle \{G*\xi\}^2(t)\rangle&=&\frac{m \,k_BT}{\omega^2}\,[1-S^2(t)]-m\,k_BT\,G^2(t),\nonumber\\
\langle \{R*\xi\}^2(t)\rangle&=&m \,k_BT\left[1-R^2(t)-\omega^2\,G^2(t)\right].
\label{conv2}
\end{eqnarray}
(A common trick to derive these results is to write the double integral over the square region
$(0,t)\times(0,t)$ of the $(t_1,t_2)$-space as two times the integral over the  triangle bounded by the lines $t_2=t_1$, $t_2=0$ and $t_1=t$.) 
The average product of the convolutions 
is easier to find by noticing that
\begin{eqnarray}
\langle \{G*\xi\}\,\{R*\xi\}\rangle=
\frac{1}{2}\, \frac{d}{dt}\,\langle \{G*\xi\}^2\rangle.
\end{eqnarray}
Then, recalling that $\dot S(t)=-\omega^2G(t)$ and $\dot G(t)=R(t)$, one obtains
\begin{eqnarray}
\langle \{G*\xi\}\,\{R*\xi\}\rangle=
m\, k_BT\, G(t)\left[S(t)-R(t)\right].
\label{conv3}
\end{eqnarray}
Finally, substituting (\ref{conv2}) and (\ref{conv3})  into (\ref{moments1}) yields
\begin{eqnarray}
\langle q^2(t)\rangle&=&\langle q_i^2\rangle\,S^2(t)+\langle v_i^2\rangle\, G^2(t)+
\frac{k_BT}{m\omega^2}\,\left[1-S^2(t)\right]-\frac{k_BT}{m}\,G^2(t),\nonumber\\
\langle v^2(t)\rangle&=&\langle q_i^2\rangle\, \omega^4\, G^2(t)+\langle v_i^2\rangle\, R^2(t)+
\frac{k_BT}{m}\,\left[1-R^2(t)-\omega^2\,G^2(t)\right], \nonumber\\
\langle q(t)\, v(t)\rangle&=&-\langle q_i^2\rangle\, \omega^2 G(t)\, S(t)+\langle v_i^2\rangle\,G(t)\, R(t)+\frac{k_BT}{m}\,G(t)\left[S(t)-R(t)\right].
\label{moments2}
\end{eqnarray}
One observes that  under conditions (\ref{cond_ergo}) the second moments relax at long times, for any initial conditions,
to the equilibrium values,
\begin{eqnarray}
\langle q^2(t)\rangle \to \frac{k_BT}{m\,\omega^2}, \qquad \langle v^2(t)\rangle \to \frac{k_BT}{m}, \qquad \langle q(t)\,v(t)\rangle\to 0.
\label{thermalization}
\end{eqnarray}
On the other hand, if conditions (\ref{cond_ergo}) are not satisfied, then it follows from the above relations that the oscillator
does not thermilize in the general case.

An exception is the case of equilibrium initial conditions. As one observes from  Eqs. (\ref{moments2}),  
if the oscillator  at $t=0$ is prepared in the state of equilibrium with $\langle q_i^2\rangle=k_BT/(m\omega^2)$
and $\langle v_i^2\rangle=k_BT/m$ (that can be arrange by connecting the oscillator at $t<0$ to an additional
thermal bath with no upper frequency cuttoff) then the terms with relaxation functions are canceled and
the moments keep the equilibrium values for $t>0$ regardless of whether asymptotic conditions (\ref{cond_ergo}) hold or not.

\section{Time correlations}
Let us show that the relaxation functions $G(t),R(t),S(t)$, introduced in the previous section, not only govern the time dependence of the moments 
$\langle q^2(t)\rangle$, $\langle v^2(t)\rangle$, $\langle q(t)v(t)\rangle$, but also determine 
the time correlation functions $\langle q(t)q(t')\rangle$,
$\langle v(t)v(t')\rangle$, $\langle q(t)v(t')\rangle$. The general expressions for the latter, valid for arbitrary initial conditions, can be obtained 
from Eqs. (\ref{solutions}) 
using the method of double Laplace 
transforms~\cite{DV,Goychuk}. Here we consider only the case of equilibrium initial conditions,
i.e. when the initial coordinate and velocity $(q_i,v_i)$ of the oscillator are drawn from the equilibrium ensemble with the moments
\begin{eqnarray}
\langle q_i\rangle=\langle v_i\rangle=\langle q_i v_i\rangle=0,
\quad
\langle q_i^2\rangle=k_BT/(m\omega^2), \quad
\langle v_i^2\rangle=k_BT/m.
\label{initial}
\end{eqnarray}

First consider the oscillator in an ergodic configuration. With equilibrium initial conditions the oscillator
remains in equilibrium also for $t>0$, and  the correlation functions depend on time only through the time difference. Then, for  $t,t'>0$ we get
from Eq. (\ref{solutions}),
\begin{eqnarray}
\langle q(t)\,q(t')\rangle&=&\langle q_i\,q(|t-t'|)\rangle=\frac{k_BT}{m\,\omega^2}\, S(|t-t'|),\nonumber\\
\langle v(t)\,v(t')\rangle&=&
\langle v_i \, v(|t-t'|)\rangle=
\frac{k_BT}{m}\, R(|t-t'|).
\label{time_corr}
\end{eqnarray}
Here the averaging is taken over the noise and also over the equilibrium distribution for initial values $q_i,v_i$.
Similarly, for the  cross-correlation we get from Eq. (\ref{solutions}):
\begin{eqnarray}
\langle q(t) v(t')\rangle
=\begin{cases}
    \langle q_i\,
    v(t'-t)\rangle=-\frac{k_BT}{m}\,G(t'-t), & \text{if $t'>t$},\\
  \langle v_i q(t-t')\rangle=\frac{k_BT}{m}\,G(t-t'), & \text{ if $t>t'$}.
  \end{cases}
\end{eqnarray}
Next, from Eqs. (\ref{solutions})  and (\ref{RH}) one can observe that the relaxation function $G(t)$ must be odd, and $R(t)$ and $S(t)$ are both even.
Then the above results can be written as
\begin{eqnarray}
\langle q(t)\,q(t')\rangle=\frac{k_BT}{m\,\omega^2}\, S(t-t'),\qquad
\langle v(t)\,v(t')\rangle=
\frac{k_BT}{m}\, R(t-t'),\quad
\langle q(t) v(t')\rangle
=\frac{k_BT}{m}
\,G(t-t').
\label{correlations}
\end{eqnarray}
Thus the relaxation functions not only determine the moments of coordinate and velocity,
but also coincide with the time correlation functions in thermal equilibrium: $S(t)$ and $R(t)$
are the normalized auto-correlation functions for the coordinate and velocity, respectively, while $G(t)$ determines the cross-correlation.

Now  consider  the oscillator in a nonergodic  configuration, and the initial conditions are still equilibrium ones, satisfying Eq. (\ref{initial}).
As was noted at the end of the previous section, in that case the moments
$\langle q^2(t)\rangle$, $\langle v^2(t)\rangle$, $\langle q(t) v(t)\rangle$  do not change with time and keep their equilibrium values for $t>0$. 
This suggests that the oscillator is in equilibrium at $t>0$, and therefore the correlations can be evaluated as above and are still given by Eq. 
(\ref{correlations}). 
Alternatively, the correlations can be evaluated directly. For equilibrium initial conditions (\ref{initial}) that method again 
recovers the results (\ref{correlations}). 

Let us demonstrate that for the cross-correlation $\langle q(t) v(t')\rangle$. From Eqs. (\ref{solutions}) and (\ref{initial}) we get
\begin{eqnarray}
\langle q(t)v(t')\rangle=\frac{k_BT}{m}\,[G(t) R(t')-S(t) G(t')+X(t,t')],
\label{aux11}
\end{eqnarray}
where $X(t,t')$ is the average product of two convolutions
\begin{eqnarray}
X(t,t')&=&\frac{1}{m\,k_BT}\,\langle \{G*\xi\}(t)\,\,\{R*\xi\}(t')\rangle.
\end{eqnarray}
We can express this function in terms of the relaxation functions as follows.
Using the fluctuation-dissipation relation (\ref{fdr}), one can write $X(t,t')$ as a double convolution
\begin{eqnarray}
X(t,t')=
\{f *\!*\, g\}(t,t')
=\int_0^t \!\! d\tau
\int_0^{t'} \!\!d\tau' \,f(t-\tau, t'-\tau')\,g(\tau,\tau')
\end{eqnarray}
with 
\begin{eqnarray}
f(t,t')=G(t)R(t'), \quad g(t,t')=K(t-t').
\end{eqnarray}
Applying the double Laplace transform
\begin{eqnarray}
\mathcal{L}_2\{\cdots\}=\int_0^\infty dt \,e^{-s t} \int_0^\infty dt'\, e^{-s't'} \{\cdots\},
\end{eqnarray}
and the convolution theorem $\mathcal L_2\{f *\!*\, g\}=\mathcal L_2\{f\}\,\mathcal L_2\{g\}$, 
we get
\begin{eqnarray}
\mathcal{L}_2\, \{X(t,t')\}=
\mathcal{L}_2 \{G(t)R(t')\}\,
\mathcal{L}_2\{K(t-t')\}=
\tilde G(s)\,\tilde R(s')\,
\mathcal{L}_2\{K(t-t')\},
\end{eqnarray}
where the tilde still denotes the one variable Laplace transform.
Referring to the properties of double Laplace transforms (see, e.g., Ref.~\cite{Deb})
and  noticing that the dissipation kernel $K(t)$ is an even function one gets
\begin{eqnarray}
\mathcal{L}_2\{K(t-t')\}=\frac{\tilde K(s)+\tilde K(s')}{s+s'},
\end{eqnarray}
and therefore
\begin{eqnarray}
\mathcal{L}_2\, \{X(t,t')\}=
\tilde G(s)\,\tilde R(s')\,
\frac{\tilde K(s)+\tilde K(s')}{s+s'}.
\end{eqnarray}
Using Eq. (\ref{laplace_relations}), 
we can make here the following replacements
\begin{eqnarray}
\tilde G(s)\tilde K(s)=\tilde S(s)-s\,\tilde G(s), \quad
\tilde R(s') \tilde K(s')=1-s'\,\tilde R(s')-\omega^2\tilde G(s')
\end{eqnarray}
to get
\begin{eqnarray}
\mathcal{L}_2\, \{X(t,t')\}=
\frac{1}{s+s'}\,[\tilde S(s)\tilde R(s')+\tilde G(s)-\omega^2\tilde G(s) \tilde G(s')]-\tilde G(s)\tilde R(s').
 \end{eqnarray}
Next, using Eq. (\ref{relaxation_laplace})
 one can make another  replacements
\begin{eqnarray}
\tilde R(s')=s'\tilde G(s')\,\quad 
\omega^2 \tilde G(s)=1-s\,\tilde S(s),
\end{eqnarray}
which yields
\begin{eqnarray}
\mathcal{L}_2\, \{X(t,t')\}=
\tilde S(s)\tilde G(s')
-\tilde G(s)\tilde R(s')+\frac{\tilde G(s)-\tilde G(s')}{s+s'}.
\end{eqnarray}
Recalling that $G(t)$ is an odd function and 
referring again to the properties of double Laplace transform one 
notices that the last term in the above expression if the double Laplace transform of $G(t-t')$,
\begin{eqnarray}
\mathcal{L}_2\, \{X(t,t')\}=
\tilde S(s)\tilde G(s')
-\tilde G(s)\tilde R(s')+\mathcal{L}_2\{G(t-t')\}.
\end{eqnarray}
Therefore in the time domain $X(t,t')$ has the form
\begin{eqnarray}
X(t,t')=S(t)G(t')-G(t)R(t')+G(t-t').
\end{eqnarray}
Finally, substituting this expression into Eq. (\ref{aux11}) we obtain
\begin{eqnarray}
\langle q(t)v(t')\rangle=
\frac{k_BT}{m}\,G(t-t'),
\end{eqnarray}
which coincides with the result (\ref{correlations}) for the ergodic oscillator in equilibrium.

 In a similar manner one can derive  the results (\ref{correlations}) for auto-correlations $\langle q(t) q(t')\rangle$ and $\langle v(t) v(t')\rangle$.
 Thus the results (\ref{correlations}), connecting the relaxation and correlation functions, 
 hold for both ergodic  and nonergodic configurations, provided initial conditions are the equilibrium ones.

\section{Relaxation functions in Laplace domain}

As was shown in the previous sections, the oscillator's dynamics can be described in terms of the relaxation functions $G(t),R(t), S(t)$.
Our goal is to find that functions in explicit forms for the specific dissipation kernel $K(t)$ given
by Eq. (\ref{K}). We shall focus on finding $G(t)$, the other two functions can be found by differentiating and integrating $G(t)$, see Eq. (\ref{RH}).  

The Laplace transform of the kernel (\ref{K}) is
\begin{eqnarray}
\tilde K(s)=\frac{\mu}{2}\,\left(
\sqrt{s^2+\omega_0^2}-s\right).
\end{eqnarray}
Substituting it into Eq. (\ref{G}) yields the Laplace transform for $G(t)$,
\begin{eqnarray}
\tilde G(s)=\frac{2}{(2-\mu)\,s^2+\mu\,s\,\sqrt{s^2+\omega_0^2} +2\,\lambda\,\omega_0^2},
\label{G_transform}
\end{eqnarray}
where we introduce the parameter  
\begin{eqnarray}
\lambda=(\omega/\omega_0)^2
\label{lambda}
\end{eqnarray}
as a dimensionless alias for the square of the oscillator frequency $\omega$.
In what follows,  we shall use $\lambda$ and $\omega$ concurrently.

The inversion of transform (\ref{G_transform}) can be expressed in terms of standard functions only for
a few special cases, see the next section. For arbitrary values of $\lambda$ and $\mu$, the inversion must be performed  by
evaluating the Bromwich integral
\begin{eqnarray}
G(t)=\frac{1}{2\pi i}\,\int_{\gamma-i\infty}^{\gamma+i\infty} e^{st} \, \tilde G(s)\,ds
\label{Bromwich_0}
\end{eqnarray}
in the complex plane. 
The character of relaxation (thermalizing  or non-thermalizing)  is determined by analytical properties of $\tilde G(s)$.
This can be anticipated as follows.
Suppose transform $\tilde G(s)$, in addition to the branch points at $\pm i\omega_0$, also has  two  simple conjugated poles
on the imaginary axis  $\pm i\omega_*$ with $\omega_*>\omega_0$. Evaluating  the Bromwich integral
by closing the integration contour, see the right part of Fig. 1, and using Cauchy's residue theorem one expects to get
contributions $e^{\pm i\omega_* t}$ which may result in the oscillatory behavior $G(t)\sim \sin\omega_* t$.
In that case the condition of thermalization (\ref{cond_ergo}) is not satisfied.
On the other hand, if $\tilde G(s)$ has no poles, or has poles on the real axis, then one expects the relaxation to thermal equilibrium.

According to Eq. (\ref{G_transform}), the pole positions  must be solutions of the equation
\begin{eqnarray}
(2-\mu)\,s^2+\mu\,s\,\sqrt{s^2+\omega_0^2} +2\,\lambda\,\omega_0^2=0.
\label{eq_poles}
\end{eqnarray}
The above consideration suggests  that the necessary condition of nonergodicity is that 
Eq. (\ref{eq_poles}) has purely imaginary solutions. The subtlety is that the function  $f(s)=\sqrt{s^2+\omega_0^2}$,
and therefore $\tilde G(s)$, has two branches and only one of the branches is physically meaningful.
Therefore special care is needed to identify the poles of 
the physical branch of $\tilde G(s)$ and to discard the poles for the unphysical branch.

\section{Special cases}
Let us first consider two cases when the inversion of the transform $\tilde G(s)$ given by Eq. (\ref{G_transform})
is known in closed analytical form. In  both cases the oscillator's  configuration is ergodic, 
so from the perspective of this paper those cases are not of particular interest.
Yet the special cases can be useful as reference points to verify the validity of the general results.

For the  first case 
\begin{eqnarray}
\mu=1,\qquad \lambda=(\omega/\omega_0)^2=1/2,
\end{eqnarray}
and transform (\ref{G_transform}) has the form 
\begin{eqnarray}
\tilde G(s)=\frac{2}{s^2+s\,\sqrt{s^2+\omega_0^2}+\omega_0^2}.
\label{G_transformA}
\end{eqnarray}
The inverse transform of this expression is given by the Bessel function,
\begin{eqnarray}
G(t)=\frac{2}{\omega_0}\, J_1(\omega_0 t).
\end{eqnarray}
The other two relaxation functions, given by Eq. (\ref{RH}), are
\begin{eqnarray}
R(t)=J_0(\omega_0 t) -J_2(\omega_0 t),\quad
S(t)=J_0(\omega_0 t).
\end{eqnarray}
All three relaxation functions vanish at long times and thus satisfy the condition of ergodic relaxation to thermal equilibrium (\ref{cond_ergo}).

The second special case corresponds to the parameter values 
\begin{eqnarray}
\mu=1,\qquad \lambda=(\omega/\omega_0)^2=1/4
\end{eqnarray}
when
\begin{eqnarray}
\tilde G(s)=\frac{4}{2\,s^2+2\,s\,\sqrt{s^2+\omega_0^2}+\omega_0^2}=\frac{4}{
\left(
s+\sqrt{s^2+\omega_0^2}
\right)^2
}.
\label{G1_B}
\end{eqnarray}
The inverse transform of this expression is known to be
\begin{eqnarray}
G(t)=\frac{8}{\omega_0^2 t}\, J_2(\omega_0 t),
\end{eqnarray}
and the other relaxation functions, according to Eq. (\ref{RH}),  are
\begin{eqnarray}
R(t)=\frac{8}{\omega_0 t}\,J_1(\omega_0 t) -\frac{24}{(\omega_0 t)^2}\,J_2(\omega_0 t),\quad
S(t)=\frac{2}{\omega_0 t}\, J_1(\omega_0 t).
\end{eqnarray}
Again, the relaxation functions describe the  ergodic relaxation since  $G(t), R(t), S(t)\to 0$ as $t\to\infty$. 
It will be shown below that for $\mu<2$ the condition of nonergodic relaxation reads  $\lambda>\lambda_c=1-\mu/2$.
For both special cases considered in this section $\lambda_c=1/2$, and the condition is not satisfied.

\section{Equations for poles}
As was noted in  Sec. V, the character of the oscillator's relaxation is governed by analytical properties of
the function $\tilde G(s)$ given by Eq. (\ref{G_transform}). That function has two branch points at $\pm i\omega_0$ and possibly a number of poles.
The positions of poles must satisfy Eq. (\ref{eq_poles}) which we rewrite here as
\begin{eqnarray}
(2-\mu)\,s^2+2 \lambda\,\omega_0^2=-\mu\,s\,f(s),
\label{eq_poles1}
\end{eqnarray}
where the function 
\begin{eqnarray}
f(s)=\sqrt{s^2+\omega_0^2}=\sqrt{(s-i\omega_0)(s+i\omega_0)},
\label{f0}
\end{eqnarray}
has two branches, which we denote as $f_1(s)$ and $f_2(s)$. Only one of the branches is physically relevant,
and our immediate goal is to define and present it in a form convenient for calculations.

\begin{figure}[t]
\includegraphics[height=5.5cm]{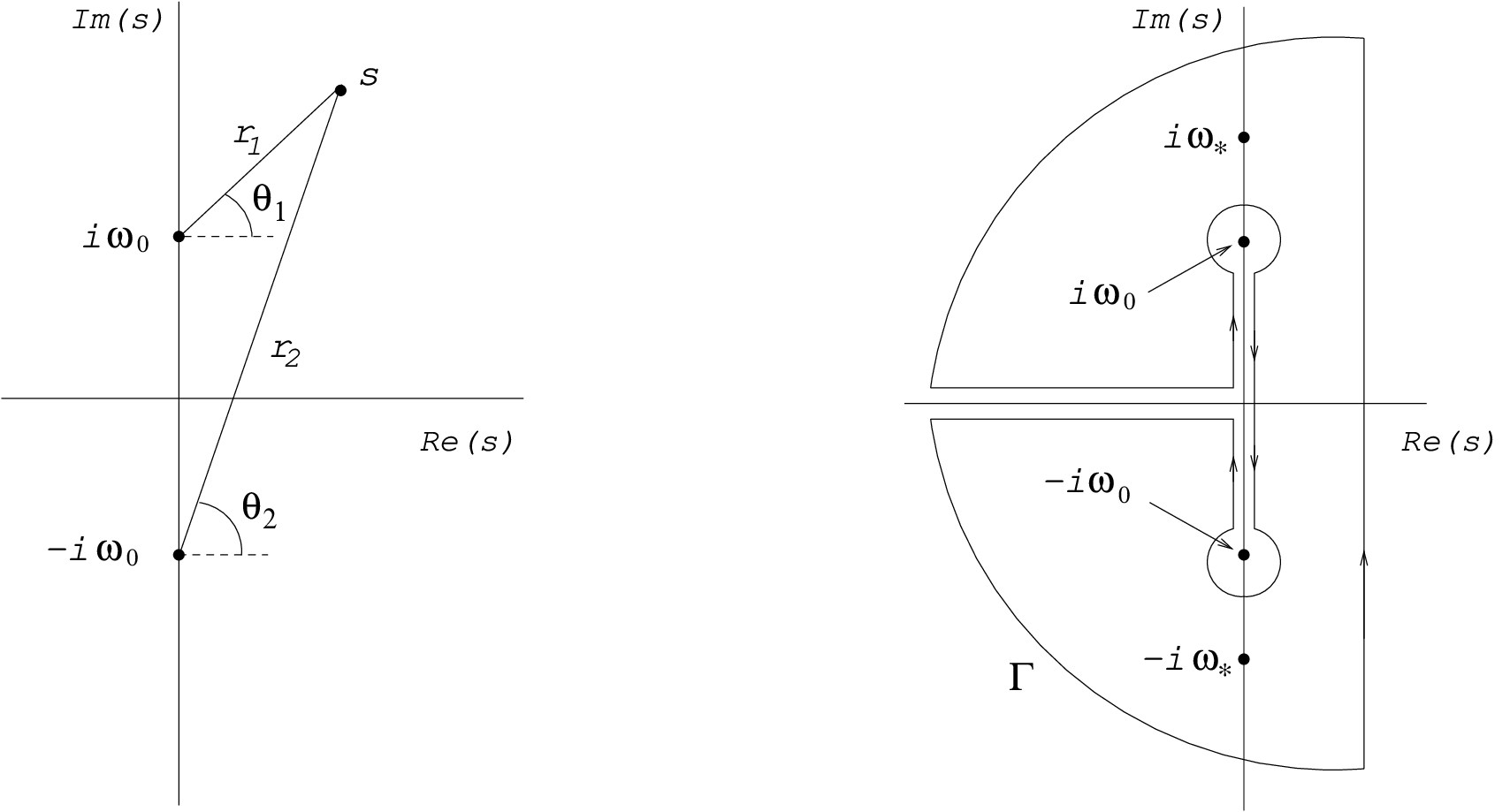}
\caption{ Left: Polar coordinates used in Eq. (\ref{polar}) to define two branches of the function $f(s)=\sqrt{s^2+\omega_0^2}$.
  Right: The integration contour $\Gamma$ for integral (\ref{I_aux}). The poles at $\pm i\omega_*$ exist only for nonergodic configurations with $\omega\ge \omega_c$.
}
\end{figure}

To this end, let us write the factors $s\pm i \omega_0$ in Eq. (\ref{f0}) in terms of polar coordinates $(r_1,\theta_1)$ and $(r_2,\theta_2)$, 
\begin{eqnarray}
s-i\omega_0=r_1\, e^{i\theta_1}, \qquad s+i\omega_0=r_2\,e^{i\theta_2},
\label{polar}
\end{eqnarray}
see the left part of Fig. 1.
Then $f(s)$ takes the form
\begin{eqnarray}
f(s)=\sqrt{r_1 r_2} \,e^{i\frac{\theta_1+\theta_2}{2}}.
\label{f}
\end{eqnarray}
Let us define the first branch $f_1(s)$ of $f(s)$ by Eq. (\ref{f}) with both polar angles in the interval $(-3\pi/2, \pi/2]$,
\begin{eqnarray}
-\frac{3\pi}{2}<\theta_1\le\frac{\pi}{2},\qquad
-\frac{3\pi}{2}<\theta_2\le\frac{\pi}{2}.
\label{f1}
\end{eqnarray}
The second  branch $f_2(s)$ is defined by Eq. (\ref{f}) with the  interval for $\theta_2$ shifted by $2\pi$,
\begin{eqnarray}
-\frac{3\pi}{2}<\theta_1\le\frac{\pi}{2},\qquad
\frac{\pi}{2}<\theta_2\le\frac{5\pi}{2}.
\label{f2}
\end{eqnarray}
The first branch $f_1(s)$ has, in particular,  the following mapping properties:

(a) If $s=x$ is real and positive (negative), then $f_1(s)$ is also real and positive (negative);

(b) If $s=\pm i \omega_*$ with real $\omega_*\ge\omega_0$, then
$f_1(s)=\pm i z$ with real $z\ge 0$.



\noindent
One can  
further elaborate  property (b): Suppose $s=\pm i\omega_*$ with $\omega_*\ge\omega_0$, then 
\begin{eqnarray}
f_1(\pm i \omega_*)=\pm i\sqrt{\omega_*^2-\omega_0^2},
\label{aux1}
\end{eqnarray}
where the square root is   the unique  positive root of a positive real number.

 As follows from Eqs.(\ref{f})-(\ref{f2}), for any $s$ the two branches of $f(s)$ are connected by the relation 
$f_2(s)=e^{i\pi} f_1(s)=-f_1(s)$.
Therefore, the mapping properties of  
the second branch $f_2(s)$ are algebraically opposite to that of the first branch, namely:

(a) If $s=x$ is real and positive (negative), then $f_2(s)$ is real and negative (positive);

(b) If $s=\pm i \omega_*$ with real $\omega_*\ge\omega_0$, then
$f_2(s)=\mp i z$ with real $z\ge 0$.


Keeping in mind the  mapping properties of $f(s)$, one can verify that
the physically meaningful branches of $\tilde K(s)$ and $\tilde G(s)$ must involve the first branch $f_1(s)$, because only in that case 
one recovers 
the correct initial conditions for $K(t)$ and $G(t)$, 
\begin{eqnarray}
K(0)=\lim_{s\to\infty} s\,\tilde K(s)=\frac{\mu\,\omega_0^2}{4},\qquad 
G(0)=\lim_{s\to\infty} s\,\tilde G(s)=0.
\end{eqnarray}
Therefore, in the equation for  poles (\ref{eq_poles1}) one has to replace  the function $f(s)$ by 
its first (physical) branch,
\begin{eqnarray}
(2-\mu)\,s^2+2 \lambda\,\omega_0^2=-\mu\,s\,f_1(s).
\label{eq_poles2}
\end{eqnarray}
Note that symbolic calculation systems like {\it Wolfram Mathematica} by default  evaluate the function $f(s)$ using its first branch.

Squaring both sides of Eq. (\ref{eq_poles2}) and moving all terms to the left hand side, one gets
\begin{eqnarray}
4(1-\mu)\,s^4+[4\lambda(2-\mu)-\mu^2]\,\omega_0^2\,s^2+4\lambda^2\omega_0^4=0.
\label{eq_poles_square}
\end{eqnarray}
Each solution of Eq. (\ref{eq_poles2}) for poles is also a solution of the squared equation  (\ref{eq_poles_square}),
but not vice versa. In other words,  positions of the poles of $\tilde G(s)$ must be among solutions
of  the squared equation (\ref{eq_poles_square}), but not every solution of Eq. (\ref{eq_poles_square}) determines a pole  of $\tilde G(s)$. 
A detailed analysis of analytical properties of $\tilde G(s)$ is somewhat different for different ranges
of $\mu$; below we shall consider those ranges separately  starting with the simpler case $\mu=1$.

\section{Poles for  $\mu=1$}
For $\mu=1$ equation (\ref{eq_poles2}) for the poles of $\tilde G(s)$ takes the form
\begin{eqnarray}
s^2+2 \lambda\,\omega_0^2=-s\,f_1(s),
\label{A1}
\end{eqnarray}
while the squared equation 
(\ref{eq_poles_square}), in general of order  four,  is reduced to a quadratic equation
\begin{eqnarray}
(4\lambda-1)\,\omega_0^2\,s^2+4\lambda^2\omega_0^4=0.
\label{B1}
\end{eqnarray}
First consider the case  $\lambda>1/4$ when Eq. (\ref{B1}) 
has two imaginary solutions
\begin{eqnarray}
s_{1,2}=\pm i \omega_*,\quad
\omega_*=\frac{2\lambda}{\sqrt{4\lambda-1}}\,\omega_0\ge \omega_0.
\label{C1}
\end{eqnarray}
Let us show that $s_{1,2}$
are also solutions of Eq. (\ref{A1}), i.e.  $\tilde G(s)$ has the poles at $s_{1,2}$, but only under the additional constraint $\lambda\ge 1/2$.

To prove that the roots $s_{1,2}$ of the squared equation (\ref{B1}) also satisfy Eq. (\ref{A1})  
it is sufficient to show that 
the left- and right-hand sides of Eq. (\ref{A1}) for $s=s_{1,2}$ have consistent signs. 
 Substituting $s_1=i\omega_*$ into Eq. (\ref{A1}) one gets for the left-hand side
\begin{eqnarray}
\text{l.h.s.}=\frac{2\lambda\,(2\lambda-1)}{4\lambda-1}\omega_0^2.
\label{lhs}
\end{eqnarray}
This expression is non-negative for $\lambda\ge 1/2$.
On the other hand, the right-hand side of Eq. (\ref{A1}) for $s=s_1$ is
always non-negative,
\begin{eqnarray}
\text{r.h.s.}=-i\omega_* f_1(i\omega_*)\ge 0.
\end{eqnarray}
This follows from the mapping property (b) for $f_1(s)$ mentioned in Sec. IV, or from Eq. (\ref{aux1}).
Therefore, the signs of the left- and right-hand sides of Eq. (\ref{A1}) for  $s=s_1$ are consistent only for $\lambda\ge 1/2$.  
A similar consideration applies for $s=s_2$. Thus we conclude that  $s_{1,2}$ given by Eq. (\ref{C1})
are solutions of the equation for poles (\ref{A1}) provided $\lambda\ge 1/2$.

The same conclusion can be arrive at by the direct evaluation of the right-hand side of Eq. (\ref{A1}) for $s=s_{1,2}$ taking into account 
Eqs. (\ref{aux1}) and (\ref{C1}). For instance, for $s=s_1$  one gets 
\begin{eqnarray}
\text{r.h.s.}=-i\omega_* f_1(i\omega_*)=\omega_*\sqrt{\omega_*^2-\omega_0^2}=
\frac{2\lambda\,|2\lambda-1|}{4\lambda-1}\,\omega_0^2.
\label{rhs}
\end{eqnarray}
For  $\lambda\ge 1/2$ this expression equals to 
the left-hand side (\ref{lhs}).   This proves that 
$s_1$, and by similar argument $s_2$, are  
solutions of Eq. (\ref{A1}) for poles under condition $\lambda\ge 1/2$.

Next consider the case $\lambda<1/4$ when  the squared equation (\ref{B1}) has
two real solutions
\begin{eqnarray}
s_{3,4}=\pm\frac{2\lambda}{\sqrt{1-4\lambda}}\,\omega_0.
\label{D1}
\end{eqnarray}
One observes that $s_{3,4}$ are not solutions of Eq. (\ref{A1}) for poles. Indeed, for $s=s_{3,4}$
the left-hand side of Eq. (\ref{A1}) is still given by expression (\ref{lhs}) which
is positive for $\lambda<1/4$. On the other hand, recalling the mapping property (a) for $f_1(s)$, see Sec. VII,
one finds that the right-hand side of Eq. (\ref{A1}) is negative for any real $s$, including  $s=s_{3,4}$.
Thus, $s_{3,4}$ do not satisfy
 Eq. (\ref{A1}) and therefore the physical branch of $\tilde G(s)$ has no poles at $s_{3,4}$.

For the remaining case $\lambda=1/4$ the equation (\ref{A1})  for poles takes the factorized form 
\begin{eqnarray}
s^2+s\,\sqrt{s^2+\omega_0^2}+\frac{\omega_0^2}{2}=\frac{1}{2}\,
\left(s+\sqrt{s^2+\omega_0^2}\right)^2=0,
\end{eqnarray}
which has no solutions.

{\it Summarizing}, for $\mu=1$   the  physical branch of $\tilde G(s)$ has the poles at $s_{1,2}=\pm i\omega_*$  under the condition 
\begin{eqnarray}
\lambda\ge\lambda_c=1/2,
\end{eqnarray}
i.e. for the oscillator frequency $\omega\ge \omega_c=\omega_0/\sqrt{2}$. The poles are located on the
the imaginary axis and for $\lambda>\lambda_c$ the corresponding frequency $\omega_*$, given by Eq. (\ref{C1}), is outside 
the bath spectrum, ($\omega_*>\omega_0$). As was discussed above and will be shown explicitly below,
under these  conditions the relaxation is nonergodic. For $\lambda<\lambda_c$
the function $\tilde G(s)$ has no poles but only the branch points at $\pm i\omega_0$. In that case
the relaxation is expected and will be shown to be ergodic. For $\lambda=\lambda_c=1/2$ we get $\omega_*=\omega_0$,
and the poles coincide with the branch points, $s_{1,2}=\pm i\omega_0$. That  is one of the special cases ($\mu=1$ and $\lambda=1/2$)
considered in Sec. VI. The relaxation was shown there to be ergodic.
Thus we conclude that for $\mu=1$ the relaxation is expected to be ergodic for $\lambda\le\lambda_c=1/2$
(for lower oscillator frequencies $\omega\le\omega_c=\omega_0/\sqrt{2}$)
and nonergodic for $\lambda>\lambda_c$ (for higher oscillator frequencies $\omega>\omega_c$).

\section{Poles for  $\mu<1$}
For $\mu\ne 1$ the squared equation  (\ref{eq_poles_square}) is of order four and has four roots which we present  as two pairs
\begin{eqnarray}
s_{1,2}=
\pm \sqrt{z_+(\lambda, \mu)}\,\omega_0, \quad 
s_{3,4}=
\pm \sqrt{z_-(\lambda, \mu)}\,\omega_0,
\label{poles_potential}
\end{eqnarray}
where 
\begin{eqnarray}
z_{\pm}(\lambda,\mu)&=&\frac{1}{8(1-\mu)}
\left(\mu^2+4\lambda(\mu-2)\pm\mu\,\sqrt{D}\right),
\label{z}
\\
D&=&16\lambda^2+8\lambda(\mu-2)+\mu^2.
\label{D}
\end{eqnarray}
We need to verify which of these roots, if any, are also solutions of the (unsquared) equation for poles (\ref{eq_poles2}).
Below we show that $\tilde G(s)$ has poles only at $s_{1,2}$, but not at $s_{3,4}$, and only 
under the condition 
\begin{eqnarray}
\lambda\ge\lambda_c=1-\mu/2.
\label{condition_x}
\end{eqnarray}

Properties of the  roots $s_{1,2}$ and $s_{3,4}$ depend on the sign of the discriminant $D$.
Consider first the case $D\ge 0$ when the functions $z_{\pm}(\lambda,\mu)$ are both real.
For the considered domain $\mu<1$, the inequality $D\ge 0$ holds when 
\begin{eqnarray}
\lambda\le\lambda_-, \quad \mbox{or}\quad \lambda\ge\lambda_+,
\end{eqnarray}
where
\begin{eqnarray}
\lambda_{\pm}=\frac{1}{4}\,\left(
2-\mu\pm 2\sqrt{1-\mu}\right)<\lambda_c.
\end{eqnarray}
One can verify that for $\lambda\le\lambda_-$ both functions $z_\pm$ are positive, so that 
all four roots (\ref{poles_potential}) are real.  But it is easy to see, recalling mapping rule (a) for $f_1(s)$ in Sec. VII,
that equation (\ref{eq_poles2}) for poles
\begin{eqnarray}
(2-\mu)\,s^2+2 \lambda\,\omega_0^2=-\mu\,s\,f_1(s).
\label{eq_poles22}
\end{eqnarray}
cannot have real solutions for the given domain $\mu<1$ since the left- and right-hand sides of the equation for real $s$
have the opposite signs. Thus we find that for $\lambda\le\lambda_-$ the function $\tilde G(s)$ has no poles.

On the other hand, for $\lambda\ge\lambda_+$ both functions 
$z_\pm$ can be shown to be negative, and all four roots (\ref{poles_potential}) are purely imaginary. Consider  
the first pair of roots, writing it as
\begin{eqnarray}
s_{1,2}=\pm i\,\omega_*, \qquad \omega_*=\beta_*(\lambda,\mu)\,\omega_0
\label{s12}
\end{eqnarray}
where the dimensionless function $\beta_*(\lambda,\mu)$ reads
\begin{eqnarray}
\beta_*(\lambda,\mu)=\sqrt{-z_+(\lambda,\mu)}=\left\{
\frac{-1}{8(1-\mu)}
\left(\mu^2+4\lambda(\mu-2)+\mu\,\sqrt{D}\right)
\right\}^{1/2}.
\label{beta*}
\end{eqnarray}
 Here the square roots are  the unique  positive roots of positive real numbers.
Let us define the critical value $\lambda_c$  for which $\beta_*(\lambda,\mu)=1$, 
\begin{eqnarray}
\beta_*(\mu,\lambda_c)
=1 \quad \Rightarrow\quad \lambda_c= 1-\mu/2.
\end{eqnarray}
Note again that $\lambda_c>\lambda_+$.
One can verify that for the considered domain $\mu<1$ the function $\beta_*(\lambda,\mu)$ for any fixed $\mu$ has a minimum at $\lambda=\lambda_c$, so that
\begin{eqnarray}
\beta_*(\lambda,\mu)\ge \beta_*(\lambda_c,\mu)=1,
\end{eqnarray}
and  the equality $\beta_*=1$ holds only for $\lambda=\lambda_c$.
Therefore, the roots $s_{1,2}$ have the structure $s_{1,2}=\pm i\omega_*$ with $\omega_*=\beta_*\omega_0\ge\omega_0$. Then 
according to Eq. (\ref{aux1}) 
\begin{eqnarray}
f_1(s_{1,2})=f_1(\pm i\omega_*)=\pm i\,\sqrt{\omega_*^2-\omega_0^2}.
\label{mapping2}
\end{eqnarray}
Taking this into account and substituting $s_{1,2}$ into equation (\ref{eq_poles22}) for poles we
find that the right-hand side of the equation is real and non-negative
\begin{eqnarray}
-\mu\,s_{1,2}\,f_1(s_{1,2})=\mu\, \omega_*\,\sqrt{\omega_*^2-\omega_0^2}\ge 
0.
\end{eqnarray}
The equation is satisfied by $s_{1,2}$ only if
the left-hand side is also non-negative,
\begin{eqnarray}
(2-\mu)\left(s_{1,2}\right)^2 +2\lambda\omega_0^2=-(2-\mu)\,\omega_*^2+2\lambda\omega_0^2\ge 
0,
\end{eqnarray}
which gives the condition 
\begin{eqnarray}
\lambda\ge\left(1-\frac{\mu}{2}
\right)\,\left(\frac{\omega_*}{\omega_0}\right)^2=\lambda_c\,\beta_*^2(\lambda,\mu).
\label{aux44}
\end{eqnarray}
Writing this as
\begin{eqnarray}
\frac{\lambda}{\lambda_c}\ge\beta_*^2(\lambda,\mu),
\label{aux55}
\end{eqnarray}
one observes that, 
since $\beta_*(\lambda,\mu)\ge 1$, the condition necessarily implies $\lambda\ge\lambda_c$. Further, one can
directly verify that the condition $\lambda\ge\lambda_c$ is not only necessary but also sufficient
for the validity of inequality (\ref{aux55}): the latter holds for {\it any} $\lambda\ge \lambda_c$. 
Thus we find that the first pair of   roots $s_{1,2}$ of the squared equation (\ref{eq_poles_square})
also satisfy the equation for poles 
(\ref{eq_poles22}), and therefore 
$\tilde G(s)$ has poles at $s_{1,2}$
under the condition $\lambda\ge\lambda_c$. The same conclusion one gets directly evaluating the left-
and right-hand sides of equation (\ref{eq_poles22}) for poles at $s=s_{1,2}$.

Consider now the second pair of roots of the squared equation (\ref{eq_poles_square}), writing them as 
\begin{eqnarray}
s_{3,4}=\pm i\,\omega_\dagger, \qquad \omega_\dagger=\beta_\dagger(\lambda,\mu)\,\omega_0
\end{eqnarray}
with 
\begin{eqnarray}
\beta_\dagger(\lambda,\mu)=
\sqrt{-z_-(\lambda,\mu)}=
\left\{
\frac{-1}{8(1-\mu)}
\left(\mu^2+4\lambda(\mu-2)-\mu\,\sqrt{D}\right)
\right\}^{1/2},
\label{beta_dagger}
\end{eqnarray}
and assuming $\lambda\ge\lambda_+$. One can verify that for the considered domain $\mu<1$ 
\begin{eqnarray}
\beta_\dagger(\lambda,\mu)> 1,\quad\text{for}\quad
\lambda\ge\lambda_+, 
\end{eqnarray}
so that  $\omega_\dagger=\beta_\dagger\,\omega_0\ge\omega_0$.
Repeating the above arguments for $s_{1,2}$ we find that the roots $s_{3,4}$ satisfy Eq. (\ref{eq_poles22}) for poles under the condition  
\begin{eqnarray}
\frac{\lambda}{\lambda_c}>\beta_\dagger^2(\lambda,\mu),
\label{aux6}
\end{eqnarray}
which is similar to condition (\ref{aux55}) for $s_{1,2}$.
One can directly verify (for instance, graphically) that inequality (\ref{aux6}) cannot be satisfied
for any $\lambda\ge \lambda_+$ and $\mu<1$. Therefore, the roots $s_{3,4}$ do not satisfy Eq. (\ref{eq_poles22})
and $\tilde G(s)$ has no poles at $s_{3,4}$.

Finally, we need to consider the interval $\lambda_-<\lambda<\lambda_+$. In that case the discriminant $D$ in Eq. (\ref{z})
is negative, and the roots $s_{1,2}$ and $s_{3,4}$
have non-zero real and imaginary parts. In that case  the simple arguments we used above,
based on the  mapping rules for the function $f(s)=\sqrt{s^2+\omega_0^2}$ for purely real or imaginary $s$, do not apply.
Yet,  an explicit evaluation (which is convenient to execute with {\it Mathematica})
shows that the imaginary parts of the left- and right-hand sides
of the equation for poles (\ref{eq_poles22}) for $s=s_{1,2}$, and 
also for $s=s_{3,4}$,  
have opposite signs. Real parts also have opposite signs except one value of $\lambda$ for which they are both zero.
Thus for  $\lambda_-<\lambda<\lambda_+$ neither $s_{1,2}$ nor $s_{3,4}$  give positions of poles of the physical branch of $\tilde G(s)$.

{\it Summarizing}, for $\mu<1$ the function  $\tilde G(s)$ 
under condition
 \begin{eqnarray}
 \lambda\ge\lambda_c=1-\mu/2
\label{cond11}
 \end{eqnarray}
  has two poles.
The poles positions are given by Eqs. (\ref{s12}) and (\ref{beta*}) and have the form $s_{1,2}=\pm i \omega_*$ with $\omega_*\ge \omega_0$.
The equality $\omega_*=\omega_0$ occurs for $\lambda=\lambda_c$; for $\lambda>\lambda_c$ the poles
  frequency $\omega_*$ is higher than the maximal mode frequency of the bath, $\omega_*>\omega_0$. Since $\lambda=(\omega/\omega_0)^2$,
  condition (\ref{cond11}) corresponds to higher values of the oscillator frequency $\omega$. For $\lambda<\lambda_c$,
  i.e.  for lower frequencies, $\tilde G(s)$ has no poles and its only singularities are the two branch points at $\pm i\omega_0$.
  These analytical properties are expected and will be shown  below to correspond to nonergodic behavior for $\lambda> \lambda_c$
  (for the higher frequency domain) and ergodic relaxation to thermal equilibrium for $\lambda\le\lambda_c$ (for the lower frequency domain).

\section{Poles for $\mu>1$}
For $\mu>1$ the roots $s_{1,2}$ and $s_{3,4}$
of
the squared equation  (\ref{eq_poles_square})
are given by the same expressions (\ref{poles_potential})-(\ref{D}) as for $\mu<1$, but now the discriminant
$D=16\lambda^2+8\lambda(\mu-2)+\mu^2$ is positive for any $\lambda$ and
the functions  $z_\pm(\lambda,\mu)$ are both real. Further one finds that
$z_+(\lambda,\mu)<0$ and $z_-(\lambda,\mu)>0$. Therefore, the first pair of roots $s_{1,2}=\pm \sqrt{z_+}$
are purely imaginary, and the second pair
$s_{3,4}=\pm \sqrt{z_-}$ are real. As was noted in the previous section, a real $s$ cannot be a solution of
Eq. (\ref{eq_poles22}) for poles. Therefore the roots  $s_{3,4}$ must be discarded, $\tilde G(s)$ has no poles there.

 Consider the purely imaginary roots $s_{1,2}$. They are given by the same expressions (\ref{s12}) and (\ref{beta*}) as for $\mu<1$,   
\begin{eqnarray}
&&s_{1,2}=\pm i\,\omega_*, \qquad \omega_*=\beta_*(\lambda,\mu)\,\omega_0,\nonumber\\
&&\beta_*(\lambda,\mu)=\sqrt{-z_+(\lambda,\mu)}=\left\{
\frac{-1}{8(1-\mu)}
\left(\mu^2+4\lambda(\mu-2)+\mu\,\sqrt{D}\right)
\right\}^{1/2}.
\label{poles555}
\end{eqnarray}
As in the previous section, let us define the critical value $\lambda_c$  for which $\beta_*(\lambda,\mu)=1$, 
\begin{eqnarray}
\beta_*(\mu,\lambda_c)
=1 \quad \Rightarrow\quad \lambda_c= 1-\mu/2.
\end{eqnarray}
In contrast to the case $\mu\le 1$ discussed in the previous sections,  a  meaningful (non-negative) $\lambda_c$ exists
only under the additional constraint $\mu\le 2$.  Thus for $\mu>1$ we need to consider separately the intervals $1<\mu<2$ and
$\mu\ge 2$.

For  $1<\mu<2$, 
substituting $s_{1,2}=\pm i \omega_*$ into Eq. (\ref{eq_poles22}) for poles, one finds as in the previous section
that the equation is satisfied under condition (\ref{aux44})
\begin{eqnarray}
\lambda\ge\lambda_c\,\beta_*^2(\lambda,\mu),
\label{aux444}
\end{eqnarray}
which holds for $\lambda\ge\lambda_c$. 

For $\mu\ge 2$, the substitution 
of $s_{1,2}=\pm i \omega_*$ into Eq. (\ref{eq_poles22}) for poles leads again to condition (\ref{aux444}),
but now  that condition  is trivially satisfied for any $\lambda>0$ since $\lambda_c=1-\mu/2\le 0$ and the right-hand side of inequality (\ref{aux444}) is non-positive.

{\it Summarizing}, for $\mu>1$  analytical properties of $\tilde G(s)$ are different for  the intervals $1<\mu<2$ and $\mu\ge 2$.
For the interval $1<\mu< 2$ we find the properties similar to that for $\mu<1$, that is  $\tilde G(s)$ has poles at $s_{1,2}$ given by Eq. (\ref{poles555})
under the condition $\lambda\ge \lambda_c=1-\mu/2$. For $\lambda=\lambda_c$ the poles coincide with the branch points at $\pm i \omega_0$.
On the other hand, for $\mu\ge 2$ the transform 
$\tilde G(s)$ has poles at $s_{1,2}$ for any value of $\lambda$.
These properties suggest the following: For $\mu< 2$ the relaxation is nonergodic for  $\lambda>\lambda_c$
(for higher oscillator frequency $\omega$) and ergodic for $\lambda\le \lambda_c$ (for lower $\omega$).
For $\mu\ge 2$ the relaxation is nonergodic for any $\lambda$ (for any $\omega$).
In what follows, these expectations will be confirmed by explicit evaluation of the relaxation functions in the time domain.

\section{Relaxation functions for $\mu=1$}

For $\mu=1$ the Laplace transform (\ref{G_transform}) of the relaxation function $G(t)$ takes the form 
\begin{eqnarray}
\tilde G(s)=\frac{2}{s^2+s\,f_1(s)+2\,\lambda\,\omega_0^2},
\label{G_transform2}
\end{eqnarray}
where $f_1(s)$ is the physical branch of the function $f(s)=\sqrt{s^2+\omega_0^2}$ defined in Sec. VII. The inverse transform is given by the 
the Bromwich integral (\ref{Bromwich_0}),
\begin{eqnarray}
G(t)=\frac{1}{2\pi i}\,\int_{\gamma-i\infty}^{\gamma+i\infty} e^{st} \, \tilde G(s)\,ds.
\label{Bromwich_00}
\end{eqnarray}
The singular points of $\tilde G(s)$ are two branch points $\pm i\omega_0$ and also
possibly two poles. 
As was discussed in Sec. VIII, 
for $\mu=1$ the poles exist 
under the condition
\begin{eqnarray}
\lambda\ge \lambda_c=1/2, \quad \mbox{or}\quad
\omega\ge \omega_c=\sqrt{1/2}\,\omega_0,
\end{eqnarray}
and have the form
\begin{eqnarray}
s_{1,2}=\pm i \omega_*,\quad
\omega_*=\frac{2\lambda}{\sqrt{4\lambda-1}}\,\omega_0\ge\omega_0.
\label{C11}
\end{eqnarray}
For $\lambda=\lambda_c$ the poles and branch points coincide,  and  for $\lambda<\lambda_c$ the function $\tilde G(s)$ has no poles.
Since all singularities are located on the imaginary axis, the integration path in Eq. (\ref{Bromwich_00}) is along a vertical line to the right of the origin, $\gamma>0$.
With the nature of singular points established, the evaluation of integral (\ref{Bromwich_00}) is a standard exercise in complex variable analysis; below we outline the main points.

The first step is to consider the auxiliary integral \begin{eqnarray}
I(t)=\frac{1}{2\pi i}\,\int_\Gamma e^{st} \, \tilde G(s)\,ds
\label{I_aux}
\end{eqnarray} 
over a closed contour $\Gamma$ shown in Fig. 1. As the radius of the arc of $\Gamma$ goes to infinity, the contribution from the arc vanishes.
The contributions  from the paths above and below the negative real axis are mutually canceled. The integrals over the small circles
around the branch points $\pm i\omega_0$ can be shown to vanish as the circles radii goes to zero.
The latter is true for any $\lambda$ including $\lambda=\lambda_c=1/2$, when the branch points coincide with the poles.
The only non-zero contributions to the integral $I(t)$ are those from the 
the rightmost vertical path and the two shores of the branch cut along the imaginary axis connecting the branch points $\pm i\omega_0$.
When the radius of the arc of $\Gamma$ goes to infinity, the contribution from the rightmost vertical path, according to Eq. (\ref{Bromwich_00}),  equals $G(t)$, therefore
\begin{eqnarray}
I(t)=G(t)+I_0(t),
\label{I2}
\end{eqnarray}
where 
\begin{eqnarray}
I_0=I_0^++I_0^-=\frac{1}{2\pi i}\,\int_{\Gamma_0^+} e^{st} \, \tilde G(s)\,ds+\frac{1}{2\pi i}\,\int_{\Gamma_0^-} e^{st} \, \tilde G(s)\,ds
\label{I0_def}
\end{eqnarray}
is the contribution  from the path along the right ($\Gamma_0^+$) and left ($\Gamma_0^-$) shores of the branch cut in the clockwise direction.
On the other hand, the integral $I(t)$ can be evaluated with Cauchy's integral and residue theorems:
\begin{eqnarray}
I(t)=\begin{cases}
    0, & \text{if $\lambda\le\lambda_c$}.\\
  \sum\limits_{i=1,2} Res[e^{st} \tilde G(s), s_i]  , & \text{ if $\lambda>\lambda_c$}.
  \end{cases}
\label{I3}
\end{eqnarray}
From (\ref{I2}) and (\ref{I3}) one gets
\begin{eqnarray}
G(t)=
\begin{cases}
-I_0(t),& \quad\mbox{for}\quad \lambda\le \lambda_c,\\
-I_0(t)+\sum\limits_{i=1,2} Res[e^{st} \tilde G(s), s_i],& \quad
\mbox{for}\quad \lambda>\lambda_c.
\end{cases}
\label{baux00}
\end{eqnarray}
The integral $I_0(t)$ along the branch cut for arbitrary $\mu$ is evaluated in Appendix A;
for $\mu=1$ the result takes the form
\begin{eqnarray}
I_0(t)=-\frac{4}{\pi\omega_0}\,\int_{0}^{1}
\frac{x\,\sqrt{1-x^2}\,\sin (x\,\omega_0 t)\,dx}{
(1-4\lambda)\, x^2+4\lambda^2}.
\label{baux0}
\end{eqnarray}
The residues 
$Res[e^{st} \tilde G(s), s_{1,2}]$ are evaluated for arbitrary $\mu$ in Appendix B; for $\mu=1$ we get
\begin{eqnarray}
Res[e^{st} \tilde G(s), s_1]+
Res[e^{st} \tilde G(s), s_2]
=\frac{1}{\omega_0}\,\frac{4\,\sqrt{\beta^2-1}}{(2\beta^2-1)+2\,\beta\,\sqrt{\beta^2-1}}\,\sin(\omega_* t), 
\label{baux1}
\end{eqnarray}
where
\begin{eqnarray}
\beta=\beta(\lambda)=\frac{\omega_*}{\omega_0}=\frac{2\lambda}{\sqrt{4\lambda-1}}.
\label{baux2}
\end{eqnarray}
This quantity has been denoted in the previous sections as $\beta_*$; from now on we drop the asterisk subscript
as superfluous.
From Eqs. (\ref{baux1}) and (\ref{baux2}) one gets a more explicit expression
\begin{eqnarray}
Res[e^{st} \tilde G(s), s_1]+
Res[e^{st} \tilde G(s), s_2]
=\frac{1}{\omega_0}\,\frac{8\lambda-4}{(4\lambda-1)^{3/2}}\,\sin(\omega_* t).
\label{baux3}
\end{eqnarray}
Finally, substituting Eqs.  (\ref{baux0}) and (\ref{baux3}) into Eq. (\ref{baux00}) yields
\begin{eqnarray}
G(t)=
\begin{cases}
    G_e(t), & \text{if $\lambda\le\lambda_c$},\\
G_e(t)+G_0\,\sin(\omega_* t), & \text{ if $\lambda>\lambda_c$}.
  \end{cases}
\label{G1}
\end{eqnarray}
where 
\begin{eqnarray}
G_e(t)=\frac{4}{\pi\omega_0}\int_0^1 \frac{\sin(x\,\omega_0\,t)\,x\,\sqrt{1-x^2}\,dx}
{(1-4\lambda)\,x^2+4\lambda^2},\quad
G_0=\frac{1}{\omega_0}\,\frac{8\,\lambda-4}{(4\,\lambda-1)^{3/2}},
\label{Ge1}
\end{eqnarray}
$\lambda_c=1/2$, and the frequency of the oscillating (nonergodic) term is $\omega_*=\beta\,\omega_0$  with $\beta=\beta(\lambda)$ given by Eq. (\ref{baux2}).  

The function $G_e(t)$ for any $\lambda$ vanishes at long times and thus represents the ergodic component of $G(t)$ (hence the subscript $e$),
while  $G_0$ is the amplitude of the nonergodic component. At long times $G(t)$ has the asymptotic form
\begin{eqnarray}
G(t)\to
\begin{cases}
    0, & \text{if $\lambda\le\lambda_c$},\\
G_0\,\sin(\omega_* t), & \text{ if $\lambda>\lambda_c$}.
  \end{cases}
\label{G1_asymp}
\end{eqnarray}
As was discussed in Sec. III, see Eq. (\ref{cond_ergo}), the asymptotic long time  condition $G(t)\to 0$ corresponds to ergodic relaxation.  
Thus, as anticipated,  the results (\ref{G1}) and (\ref{G1_asymp}) show that the oscillator is ergodic
(reaches thermal equilibrium with the bath at long times) when  $\lambda\le\lambda_c=1/2$.  For  $\lambda>\lambda_c$, 
the time-periodic  component of $G(t)$ develops; the oscillator is nonergodic and does not thermalize.

\begin{figure}[t]
\includegraphics[height=6.0cm]{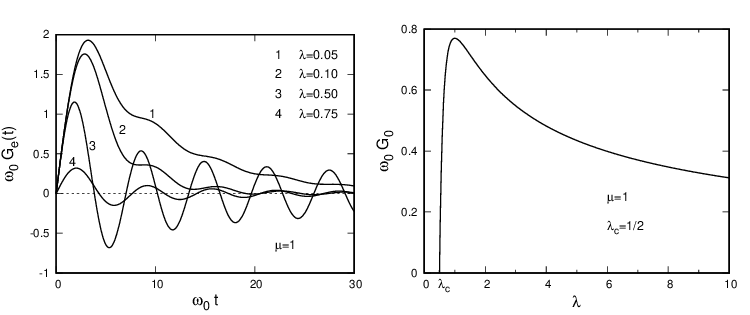}
\caption{Left: The ergodic component $G_e(t)$ of the relaxation function $G(t)$, see Eqs. (\ref{G1}) and (\ref{Ge1}),
  for the mass ratio parameter $\mu=1$ and several values of the oscillator frequency parameter $\lambda=(\omega/\omega_0)^2$.
  Right: The amplitude $G_0$ of the  nonergodic time-periodic component of $G(t)$ as a function of $\lambda$.
  It is zero for $\lambda\le\lambda_c=1/2$ and has a maximum at $\lambda=1$.
}
\end{figure}

For $\lambda=1/4$ and $\lambda=1/2$ 
the integral form of 
the ergodic component $G_e(t)$ given by Eq. (\ref{Ge1})
can be expressed 
in terms of  the Bessel functions,  
\begin{eqnarray}
G(t)=G_e(t)=\begin{cases}
    \frac{8}{\omega_0^2 t}\, J_2(\omega_0 t), & \text{for $\lambda=1/4$},\\
    \frac{2}{\omega_0}\,J_1(\omega_0 t), & \text{for $\lambda=1/2$.}
\end{cases}
\end{eqnarray}
Those are two special solutions already found in Sec. VI using a table of standard Laplace transforms.

For several values of $\lambda<1$ the ergodic component $G_e(t)$ is presented on the left plot of  Fig. 2. For $\lambda>1$, $G_e(t)$
has an oscillatory decaying shape similar to that for  $\lambda=1/2$,
but its range quickly decreases with increasing $\lambda$; for instance, for $\lambda=5$ 
the maximum value of $G_e(t)$ is
of order of $10^{-3}$.

The amplitude $G_0(\lambda)$
of the nonergodic oscillatory component, given by Eq. (\ref{Ge1}),
as a function of $\lambda$
is shown on the right plot of Fig. 2.
It is zero for $\lambda\le\lambda_c$, while for $\lambda>\lambda_c$ it first quickly increases, reaches a maximum at $\lambda=1$,
and then monotonically decreases as $1/\sqrt{\lambda}$.

Differentiating and integrating Eq. (\ref{G1}) yield the other two relaxation functions $R(t)$ and $S(t)$, see Eq. (\ref{RH}).
They have the structure similar to $G(t)$, i.e. have only ergodic component for $\lambda\le \lambda_c$
and both ergodic and  nonergodic components for $\lambda>\lambda_c$. For $R(t)=\frac{d}{dt}G(t)$ we get
\begin{eqnarray}
R(t)=
\begin{cases}
    R_e(t), & \text{if $\lambda\le\lambda_c$},\\
R_e(t)+R_0\,\cos(\omega_* t), & \text{ if $\lambda>\lambda_c$},
  \end{cases}
\label{R1}
\end{eqnarray}
where the ergodic component $R_e(t)$ and the amplitude $R_0$ of the nonergodic component are 
\begin{eqnarray}
R_e(t)=\frac{4}{\pi}\int_0^1 \frac{\cos(x\,\omega_0\,t)\,x^2\,\sqrt{1-x^2}\,dx}
{(1-4\lambda)\,x^2+4\lambda^2},\quad
R_0=\beta\omega_0 G_0=\frac{8\,\lambda(2\lambda-1)}{(4\,\lambda-1)^2}.
\label{Re1}
\end{eqnarray}
The functions $R_e(t)$ and $R_0(\lambda)$ are presented in Fig. 3.
Similar to $G_e(t)$, $R_e(t)$ vanishes  at long times for any $\lambda$, and thus can be interpreted as an ergodic component of $R(t)$. 
Note that $R_e(0)=1$ for $\lambda\le \lambda_c$ and $R_e(0)+R_0=1$ for $\lambda>\lambda_c$, so  that $R(0)=1$ for any $\lambda$.
This is the correct initial condition which can be found  without  inverting  $\tilde R(s)$, see Eq. (\ref{GHR0}).

\begin{figure}[t]
\includegraphics[height=6.0cm]{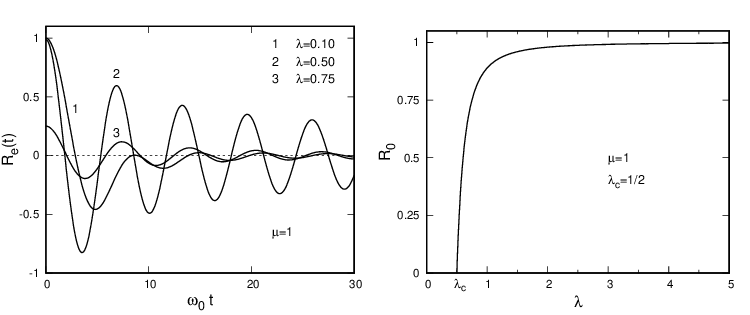}
\caption{Left: The ergodic component $R_e(t)$ of the relaxation function $R(t)$, see Eqs. (\ref{R1}) and (\ref{Re1}),
  for the mass ratio parameter $\mu=1$ and several values of the frequency parameter $\lambda=(\omega/\omega_0)^2$.
  Right: The amplitude $R_0$ of the  nonergodic time-periodic component of $R(t)$ as a function of $\lambda$.
  The nonergodic component is zero for $\lambda\le\lambda_c=1/2$.
}
\end{figure}

For the relaxation function $S(t)=1-\omega^2\int_0^t G(t) dt$
we get
\begin{eqnarray}
S(t)=
\begin{cases}
    S_e(t), & \text{if $\lambda\le\lambda_c$},\\
S_e(t)+S_0\,[\cos(\omega_* t)-1], & \text{ if $\lambda>\lambda_c$}.
  \end{cases}
\label{H1}
\end{eqnarray}
with
\begin{eqnarray}
S_e(t)=
1-\frac{4\lambda}{\pi}\int_0^1 \frac{[1-\cos(x\,\omega_0\,t)]\,\sqrt{1-x^2}\,dx}
{(1-4\lambda)\,x^2+4\lambda^2},\quad
S_0=\frac{\lambda\omega_0 G_0}{\beta}=\frac{4\,\lambda-2}{4\,\lambda-1}.
\label{He1}
\end{eqnarray}
The functions $S_e(t)$ and $S_0(\lambda)$ are presented in Fig. 4.
At long times
$S_e(t)$ has the asymptotic time-independent form
\begin{eqnarray}
S_e(t)\to
1-\frac{4\lambda}{\pi}\int_0^1 \frac{\sqrt{1-x^2}\,dx}
{(1-4\lambda)\,x^2+4\lambda^2},
\end{eqnarray}
which takes different values  for $\lambda\le \lambda_c$ and for $\lambda> \lambda_c$, namely
\begin{eqnarray}
S_e(t)\to
\begin{cases}
    0, & \text{if $\lambda\le\lambda_c$},\\
S_0, & \text{ if $\lambda>\lambda_c$}.
  \end{cases}
\label{He_asymp}
\end{eqnarray}
As the result,  
similar to the other two relaxation functions, $S(t)$ at long times
vanishes for ergodic configurations
and oscillates about zero for nonergodic ones,
\begin{eqnarray}
S(t)\to
\begin{cases}
    0, & \text{if $\lambda\le\lambda_c$},\\
S_0\,\cos\omega_* t, & \text{ if $\lambda>\lambda_c$}.
  \end{cases}
\label{H_asymp}
\end{eqnarray}

{\it Summarizing}, in this section we obtained explicit expressions for the relaxation functions for the case $\mu=1$.
For $\lambda\le\lambda_c$ (for lower values of the oscillator frequency,   $\omega\le\omega_c=\sqrt{\lambda_c}\,\omega_0=\omega_0/\sqrt{2}$),
the relaxation functions  vanish at long times.
As was discussed in Sec. III, such behavior corresponds to the ergodic relaxation to thermal equilibrium.
On the other hand, for $\lambda>\lambda_c$ (for higher frequencies $\omega>\omega_c$) the relaxation functions develop time-periodic terms
which do not vanish at long times but oscillate about zero. As a result, the oscillator does not thermalize
but reaches a cyclostationary non-equilibrium state characterized by the oscillatory behavior of the relaxation functions.
Below we show that similar results hold not only for $\mu=1$ but for the entire domain $\mu<2$.

\begin{figure}[t]
\includegraphics[height=6.0cm]{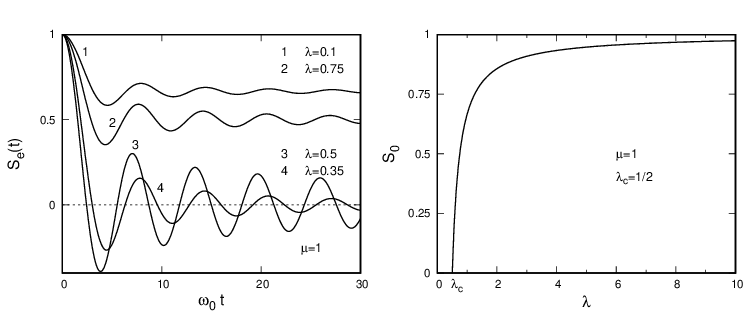}
\caption{Left: The ergodic component $S_e(t)$ of the relaxation function $S(t)$, see Eqs. (\ref{H1}) and (\ref{He1}),
  for the mass ratio parameter $\mu=1$ and several values of the frequency parameter $\lambda=(\omega/\omega_0)^2$.
Right: The amplitude $S_0$ of the  nonergodic component of $S(t)$ as a function of $\lambda$.  
}
\end{figure}

\section{Relaxation functions for $\mu<2$} 
In Secs.  IX and X we found that for the intervals $\mu<1$ and $1<\mu<2$, and also  
under the condition 
\begin{eqnarray}
\lambda\ge \lambda_c=1-\mu/2, \quad\mbox{or}\quad
\omega \ge\omega_c=\sqrt{1-\mu/2}\,\,\omega_0
\end{eqnarray}
the Laplace transform $\tilde G(s)$ of the relaxation function $G(t)$ 
has two poles on the imaginary axis. The poles are given by the following expressions: 
\begin{eqnarray}
&&s_{1,2}=\pm i\,\omega_*, \qquad \omega_*=\beta(\lambda,\mu)\,\omega_0,\nonumber\\
&&\beta(\lambda,\mu)=\left\{
\frac{1}{8(\mu-1)}
\left(
\mu^2+4\lambda(\mu-2)+\mu\,\sqrt{16\lambda^2+8\lambda(\mu-2)+\mu^2}
\right)
\right\}^{1/2},
\label{poles5555}
\end{eqnarray}
and $\lambda_c$ is defined by the equation $\beta(\lambda_c,\mu)=1$.
On the other hand, for $\lambda<\lambda_c$ the transform $\tilde G(s)$ has no poles.
To find $G(t)$ for the combined domain
\begin{eqnarray}
\mu\in(0,1)\cup (1,2)
\label{domain_B}
\end{eqnarray}
by inversion of  $\tilde G(s)$ 
we follow the same procedure  as in the previous section to find again  expression (\ref{baux00}), 
\begin{eqnarray}
G(t)=
\begin{cases}
-I_0(t),& \quad\mbox{for}\quad \lambda\le \lambda_c,\\
-I_0(t)+\sum\limits_{i=1,2} Res[e^{st} \tilde G(s), s_i],& \quad
\mbox{for}\quad \lambda>\lambda_c,
\end{cases}
\label{daux00}
\end{eqnarray}
where the integral 
\begin{eqnarray}
I_0=I_0^++I_0^-=\frac{1}{2\pi i}\,\int_{\Gamma_0^+} e^{st} \, \tilde G(s)\,ds+\frac{1}{2\pi i}\,\int_{\Gamma_0^-} e^{st} \, \tilde G(s)\,ds
\label{I0_def2}
\end{eqnarray}
is along the right ($\Gamma_0^+$) and left ($\Gamma_0^-$) shores of
the branch cut in the clockwise direction. As shown in Appendix A, for arbitrary $\mu$ this integral has the form 
\begin{eqnarray}
I_0(t)=-\frac{4\mu}{\pi\omega_0}\,\int_{0}^{1}
\frac{x\,\sqrt{1-x^2}\,\sin (x\,\omega_0 t)\,dx}{4(1-\mu)\,x^4
+[4\lambda(\mu-2)+\mu^2]\, x^2+4\lambda^2}.
\label{I0_result2}
\end{eqnarray}
The sum of residues is evaluated in Appendix B, 
\begin{eqnarray}
Res[e^{st} \tilde G(s), s_1]+
Res[e^{st} \tilde G(s), s_2]
=\frac{1}{\omega_0}\,\frac{4\,\sqrt{\beta^2-1}}{\mu\,(2\beta^2-1)+2(2-\mu)\beta\,\sqrt{\beta^2-1}}\,\sin(\omega_* t).
\label{res_result2}
\end{eqnarray}
Substituting Eqs.(\ref{I0_result2}) and (\ref{res_result2}) into Eq.(\ref{daux00}) yields 
\begin{eqnarray}
G(t)=
\begin{cases}
    G_e(t), & \text{if $\lambda\le\lambda_c$},\\
G_e(t)+G_0\,\sin(\omega_* t), & \text{ if $\lambda>\lambda_c$}.
  \end{cases}
\label{G2}
\end{eqnarray}
where the ergodic component $G_e(t)$ and the amplitude $G_0$ of the nonergodic oscillatory term are 
\begin{eqnarray}
G_e(t)&=&\frac{4\mu}{\pi\omega_0}\,\int_{0}^{1}
\frac{x\,\sqrt{1-x^2}\,\sin (x\,\omega_0 t)\,dx}{4(1-\mu)\,x^4
+[4\lambda(\mu-2)+\mu^2]\, x^2+4\lambda^2},\nonumber\\
G_0&=&\frac{1}{\omega_0}\,\frac{4\,\sqrt{\beta^2-1}}{\mu\,(2\beta^2-1)+2(2-\mu)\beta\,\sqrt{\beta^2-1}},
\label{G22}
\end{eqnarray}
the frequency of the nonergodic term is $\omega_*=\beta\omega_0$, and $\beta$ is given by Eq. (\ref{poles5555}). 

\begin{figure}[t]
\includegraphics[height=6.0cm]{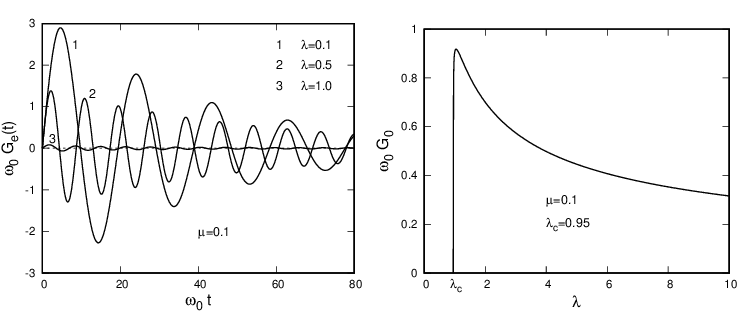}
\caption{Left: The ergodic component $G_e(t)$ of the relaxation function $G(t)$, see Eqs. (\ref{G2}) and (\ref{G22}),  for the mass ratio  parameter $\mu=0.1$ and several values of the frequency parameter $\lambda=(\omega/\omega_0)^2$.
Right: The amplitude $G_0$ of the  nonergodic component of $G(t)$ as a function of $\lambda$ for $\mu=0.1$.   The nonergodic term is zero  for $\lambda\le\lambda_c=1-\mu/2=0.95$.
}
\end{figure}

For the relaxation function 
$R(t)=dG(t)/dt$ we get
\begin{eqnarray}
R(t)&=&
\begin{cases}
    R_e(t), & \text{if $\lambda\le\lambda_c$},\\
R_e(t)+R_0\,\cos(\omega_* t), & \text{ if $\lambda>\lambda_c$},\nonumber
  \end{cases}\\
R_e(t)&=&\frac{4\mu}{\pi}\,\int_{0}^{1}
\frac{x^2\,\sqrt{1-x^2}\,\cos (x\,\omega_0 t)\,dx}{4(1-\mu)\,x^4
+[4\lambda(\mu-2)+\mu^2]\, x^2+4\lambda^2},\nonumber\\
R_0&=&\beta\omega_0G_0=\frac{4\,\beta\,\sqrt{\beta^2-1}}{\mu\,(2\beta^2-1)+2(2-\mu)\beta\,\sqrt{\beta^2-1}},
\label{R2}
\end{eqnarray}

\begin{figure}[t]
\includegraphics[height=6.0cm]{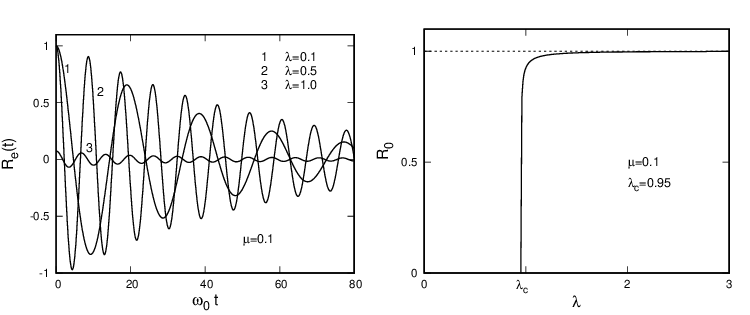}
\caption{Left: The ergodic component $R_e(t)$ of the relaxation function $R(t)$, see Eq. (\ref{R2}),
  for the mass ratio  parameter $\mu=0.1$ and several values of the frequency parameter $\lambda=(\omega/\omega_0)^2$.
Right: The amplitude $R_0$ of the  nonergodic component of $R(t)$ as a function of $\lambda$ for $\mu=0.1$. 
}
\end{figure}

Finally, for the relaxation function $S(t)=1-\omega^2\int_0^t G(t) dt$ we obtain 
\begin{eqnarray}
S(t)&=&
\begin{cases}
    S_e(t), & \text{if $\lambda\le\lambda_c$},\\
S_e(t)+S_0\,[\cos(\omega_* t)-1], & \text{ if $\lambda>\lambda_c$},\nonumber
  \end{cases}\\
S_e(t)&=&1-\frac{4\lambda\mu}{\pi}\,\int_{0}^{1}
\frac{\sqrt{1-x^2}\,[1-\cos (x\,\omega_0 t)]\,dx}{4(1-\mu)\,x^4
+[4\lambda(\mu-2)+\mu^2]\, x^2+4\lambda^2},\nonumber\\
S_0&=&\frac{\lambda}{\beta}\, \omega_0 G_0
=\frac{\lambda}{\beta}\,\frac{4\,\sqrt{\beta^2-1}}{\mu\,(2\beta^2-1)+2(2-\mu)\beta\,\sqrt{\beta^2-1}}.
\label{H2}
\end{eqnarray}

At long times the time-dependent contribution in the expression for $S_e(t)$ vanishes, and $S_e(t)$ takes the asymptotic form
\begin{eqnarray}
S_e(t)\to 1-\frac{4\lambda\mu}{\pi}\,\int_{0}^{1}
\frac{\sqrt{1-x^2}\,dx}{4(1-\mu)\,x^4
+[4\lambda(\mu-2)+\mu^2]\,
x^2+4\lambda^2}.
\label{aux4}
\end{eqnarray}
One can verify that, similar to the case $\mu=1$, the asymptotic expression (\ref{aux4}) vanishes for $\lambda\le \lambda_c$ and equals $S_0$  otherwise,
\begin{eqnarray}
S_e(t)\to
\begin{cases}
    0, & \text{if $\lambda\le\lambda_c$},\\
S_0, & \text{ if $\lambda>\lambda_c$}.
  \end{cases}
\label{Se_asymp}
\end{eqnarray}
Then, as follows from Eqs.(\ref{H2}) and (\ref{Se_asymp}), $S(t)$ at long times
vanishes for ergodic configurations
and oscillates about zero for nonergodic ones,
\begin{eqnarray}
S(t)\to
\begin{cases}
    0, & \text{if $\lambda\le\lambda_c$},\\
S_0\,\cos\omega_* t, & \text{ if $\lambda>\lambda_c$}.
  \end{cases}
\label{S_asymp}
\end{eqnarray}
Two other relaxation functions have the similar asymptotic forms. 

The behavior of the ergodic  and nonergodic components of 
the relaxation  functions
$G(t), R(t),S(t)$ for $\mu=0.1$ is illustrated in Figs. 5, 6, and 7, respectively.
The behavior is qualitatively similar to that for the case $\mu=1$, discussed in the previous section.
However, the ergodic components as functions of time decay faster, and the increase of the amplitudes of
the nonergodic components (the initial increase for $G_0$) as functions of $\lambda$ is  steeper than  for $\mu=1$.

The above results for the relaxation functions hold for $\mu$ in the interval (\ref{domain_B}), i.e. for $0<\mu<2$ except $\mu=1$.  They
do not directly apply for $\mu=1$ because parameter $\beta$, given by Eq. (\ref{poles5555}),  is not defined for $\mu=1$. Yet  one observes that the limit
\begin{eqnarray}
\lim_{\mu\to 1} \beta(\lambda,\mu)=\frac{2\lambda}{\sqrt{4\lambda-1}}
\end{eqnarray}
coincides with the result  we found for $\beta=\omega_*/\omega_0$ for $\mu=1$, see Eq. (\ref{C11}).
With that value for $\beta$ and $\mu=1$, one finds that the expressions obtained in this section recover those 
we found in Sec. XI  for $\mu=1$. Therefore, 
if expression (\ref{poles5555}) for the  function 
$\beta(\lambda,\mu)$ is defined at $\mu=1$ by continuity,
\begin{eqnarray}
\beta(\lambda,\mu)=
\begin{cases}
\left\{
\frac{1}{8(\mu-1)}
\left(
\mu^2+4\lambda(\mu-2)+\mu\,\sqrt{16\lambda^2+8\lambda(\mu-2)+\mu^2}
\right)
\right\}^{1/2}, & \text{if $\mu\ne 1$},\\
\frac{2\lambda}{\sqrt{4\lambda-1}}, & \text{ if $\mu=1$},
  \end{cases}
  \label{beta_gen}
\end{eqnarray}
then the results of this sections  hold for the whole range $\mu<2$ including $\mu=1$.

{\it Summarizing} the results of this and previous sections, 
we found that for the whole interval $\mu<2$ the relaxation is ergodic for $\lambda\le\lambda_c=1-\mu/2$
(the relaxation functions vanish at long times) and nonergodic for $\lambda>\lambda_c$
(the relaxation functions at long times oscillate about zero).
The relaxation functions are given by Eqs. (\ref{G2})-(\ref{H2}), which hold for the whole interval
$\mu<2$, while the frequency of the nonergodic component is $\omega_*=\beta\,\omega_0$, where $\beta(\alpha,\mu)$
is a continuous function determined by Eq. (\ref{beta_gen}).

\begin{figure}[t]
\includegraphics[height=6.0cm]{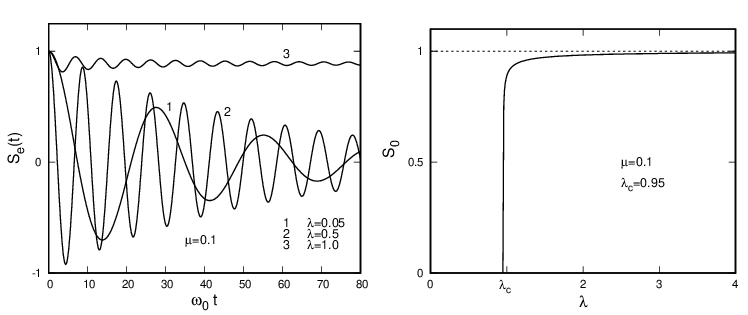}
\caption{Left: The ergodic component $S_e(t)$ of the relaxation function $S(t)$, see Eq. (\ref{H2}),
  for the mass ratio  parameter $\mu=0.1$ and several values of the  frequency parameter $\lambda=(\omega/\omega_0)^2$.
Right: The amplitude $S_0$ of the  nonergodic component of $S(t)$ as a function of $\lambda$ for $\mu=0.1$.  
}
\end{figure}

\section{Relaxation functions for $\mu\ge 2$}
In Sec. X we found that for $\mu\ge 2$  the function $\tilde G(s)$ has poles at $s=\pm i\omega_*=\pm i\beta\omega_0$
for any value of the frequency parameter $\lambda=(\omega/\omega_0)^2$. 
For that case,  we obtain
\begin{eqnarray}
G(t)&=&G_e(t)+G_0\sin\omega_* t,\nonumber\\
R(t)&=&R_e(t)+R_0\cos\omega_* t,\nonumber\\
S(t)&=&S_e(t)+S_0[\cos\omega_* t-1],
\end{eqnarray}
where the functions $G_e(t),R_e(t),S_e(t)$, the amplitudes $G_0,R_0,S_0$, and the frequency $\omega_*$
are given by expressions of the previous section. Thus, for $\mu\ge 2$ the oscillator
has only nonergodic  configurations and does not thermalize for any $\lambda$, i.e. for any value of the oscillator frequency $\omega$. 
Fig. 8 illustrates the behavior of $G_e(t)$ and $G_0(\lambda)$ for $\mu\ge 2$. Interestingly, for larger values of $\mu$
the interval of the initial increase of 
the function $G_0(\lambda)$
vanishes, and the function decreases monotonically for all $\lambda$. 

\begin{figure}[t]
\includegraphics[height=6.0cm]{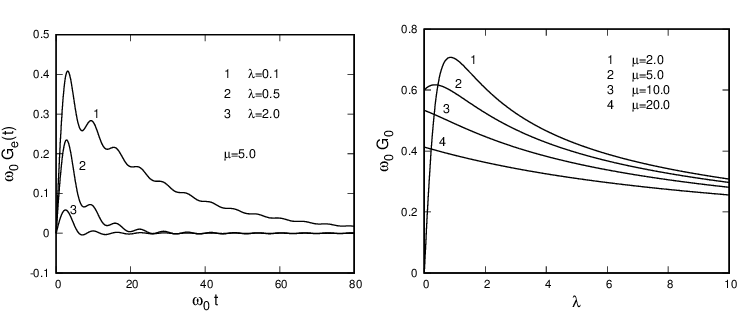}
\caption{Left: The ergodic component $G_e(t)$ of the relaxation function $G(t)$ for the mass ratio  parameter $\mu=5.0$
  and several values of the oscillator frequency parameter $\lambda=(\omega/\omega_0)^2$.
Right: The amplitude $G_0$ of the  nonergodic component of $G(t)$ as a function of $\lambda$ for several values $\mu\ge 2$.  
}
\end{figure}

\section{Ergodic to nonergodic transitions}

As an application of the results, let us 
consider a setting when the oscillator frequency $\omega$, and the frequency parameter $\lambda=(\omega/\omega_0)^2$,
can be changed instantaneously by an external agent. Suppose that the mass ratio parameter is $\mu<2$.
In that case the oscillator has both ergodic configurations corresponding to $\omega\le \omega_c$ and nonergodic
ones corresponding to   $\omega>\omega_c$, and the critical frequency is 
\begin{eqnarray}
\omega_c=\sqrt{\lambda_c}\,\omega_0=\sqrt{1-\mu/2}\,\omega_0.
\end{eqnarray}

Let assume that at $t<0$ the oscillator is in an ergodic initial configuration with the frequency  $\omega_i<\omega_c$.
Then at $t=0$ the oscillator is in thermal equilibrium with the average energy $E(\omega_i)=k_BT$
and the coordinate's variance  $\langle q_i^2\rangle=k_BT/(m\omega_i^2)$.
 At $t=0$ the frequency is instantaneously changed, $\omega_i\to\omega$. 
 If the new frequency $\omega$  is lower than or equal to $\omega_c$, then the new configuration is also ergodic, so the oscillator, 
 after some transient time, will reach again the equilibrium state with the same energy as for the  initial configuration, $E(\omega)=k_BT$.   
 We may call that process an ergodic to ergodic transition. Its characteristic feature is that, except for a transient initial relaxation, 
 the oscillator average energy does not change.
 Using an ergodic to ergodic transition, an external agent can temporarily supply the oscillator
 with a large amount of energy, but the oscillator is unable to keep it for long; in the course
 of time the energy surplus  dissipates into the bath. For an ergodic oscillator that is the only scenario.
 
 Now suppose the new frequency is higher than the critical value,  $\omega> \omega_c$. In that case 
 the oscillator does not  thermalize, but instead reaches at long times a cyclostationary state,
 characterized by  the oscillatory time dependence of the relaxation and correlation functions.
 The average energy also oscillates in time and depends on both initial $\omega_i$ and final $\omega$
 frequencies. One may say that the system undergoes an ergodic to nonergodic transition.
 Clearly the properties of such transition   depend on the protocol of switching $\omega_i\to \omega$,
 or $\lambda_i\to\lambda$. The presented results allow us to discuss only the case when the switching occurs
 instantaneously. If instead  the switching takes a finite time and is described by a smooth function $\lambda(t)$,  then
 the properties of the cyclostationary final state would be different. That case is more difficult because requires
 to solve the generalized Langevin equation with a time-dependent oscillator frequency.

The average oscillator energy after the instantaneous switching $\omega_i\to\omega$ at time $t=0$ is
\begin{eqnarray}
E(t)=\frac{m\,\omega^2}{2}\,\langle q^2(t)\rangle+
\frac{m}{2}\langle v^2(t)\rangle,
\label{energy}
\end{eqnarray} 
where the second moments of the coordinate and velocity are given by expressions (\ref{moments2}). With equilibrium initial conditions
\begin{eqnarray}
\langle q_i^2\rangle=\frac{k_BT}{m\omega_i^2}, \qquad
\langle v_i^2\rangle=\frac{k_BT}{m}
\end{eqnarray}
that expressions take the form
\begin{eqnarray}
\langle q^2(t)\rangle&=&
\frac{k_BT}{m\omega_i^2}S^2(t)
+\frac{k_BT}{m\omega^2} 
\left[1-S^2(t)\right],
\nonumber\\
\langle v^2(t)\rangle&=&
\frac{k_BT}{m}\,\left[1-\omega^2\,G^2(t)\right]+
\frac{k_BT}{m\omega_i^2}\, \omega^4\, G^2(t). 
\label{moments4}
\end{eqnarray}
From Eqs. (\ref{energy}) and (\ref{moments4}) we get 
\begin{eqnarray}
E(t)=k_BT+ \frac{k_B T}{2}\left[
\left(\frac{\omega}{\omega_i}\right)^2-1\right] 
\Big\{
S^2(t)+\omega^2G^2(t)
\Big\}.
\label{energy2}
\end{eqnarray}
In terms of $\lambda=(\omega/\omega_0)^2$ and $\lambda_i=(\omega_i/\omega_0)^2$ the result reads
\begin{eqnarray}
E(t)=k_BT+ \frac{k_B T}{2}\left[
\frac{\lambda}{\lambda_i}-1\right] 
\Big\{
S^2(t)+\lambda\,\omega_0^2G^2(t)
\Big\}.
\label{energy3}
\end{eqnarray}
For  $\lambda\le \lambda_c=1-\mu/2$, the relaxation functions have only ergodic components vanishing at long times
\begin{eqnarray}
G(t)=G_e(t)\to 0, \quad
S(t)=S_e(t)\to 0, \quad \text{as}
\quad t\to\infty
\label{responce1}
\end{eqnarray}
In that case, as expected,  Eq. (\ref{energy3}) shows that the oscillator's energy relaxes to the equilibrium value $k_BT$.

Now suppose $\lambda>\lambda_c$. In that case the relaxation functions have both ergodic and nonergodic components,
\begin{eqnarray}
G(t)=G_e(t)+G_0\,\sin\omega_* t, \quad
S(t)=S_e(t)+S_0\,[\cos(\omega_* t)-1],
\end{eqnarray}
and the ergodic components  have the asymptotic properties 
\begin{eqnarray}
G_e(t)\to 0, \quad
S_e(t)\to S_0, \quad \text{as}
\quad t\to\infty.
\label{responce2}
\end{eqnarray}
Then, taking into account that 
$S_0=\lambda\omega_0 G_0/\beta$, see Eq. (\ref{H2}),
one finds in the limit of 
long times the oscillator energy  in the cyclostationary ({\it cs}) state:
\begin{eqnarray}
E_{cs}(t)=k_BT+ \frac{k_B T}{2}\left(
\frac{\lambda}{\lambda_i}-1\right)
(\omega_0 G_0)^2
\left\{
\left(
\frac{\lambda}{\beta}
\right)^2
\,
\cos^2(\omega_* t)
+
\lambda\,\sin^2(\omega_* t)
\right\}.
\end{eqnarray}
Instead of relaxing to the equilibrium value $k_BT$, the energy oscillates with time, and its lower bound exceeds
the equilibrium value $k_BT$.
After the additional averaging over time (denoted by the overbar), the energy of the oscillator in the cyclostationary  state takes the form
\begin{eqnarray}
\overline{E}_{cs}=
k_BT+ \frac{k_B T}{4}\left(
\frac{\lambda}{\lambda_i}-1\right)
\left[
\left(\frac{\lambda}{\beta}
\right)^2
+
\lambda
\right]\,(\omega_0 G_0)^2
\end{eqnarray}
Here $\omega_0G_0$ and $\beta=\beta(\lambda,\mu)$ are given by Eqs. (\ref{G22}) and  (\ref{beta_gen}), respectively.

The physical interpretation of the above results is as follows. 
When the external agent at $t=0$  instantaneously increases the oscillator frequency 
$\omega_i\to\omega$, the  oscillator (potential) energy is increased by the amount 
\begin{eqnarray}
\Delta E=\frac{m}{2}(\omega^2-\omega_i^2)\langle q_i^2\rangle=\frac{k_BT}{2}\left[\left(
\frac{\omega}{\omega_i}\right)^2-1\right]=
\frac{k_BT}{2}\left(
\frac{\lambda}{\lambda_i}-1\right).
\end{eqnarray}
Thus at $t=0^+$ the oscillator, just kicked out of equilibrium,  has the energy 
\begin{eqnarray}
E(0^+)=k_BT+\Delta E=k_BT+\frac{k_BT}{2}\left(
\frac{\lambda}{\lambda_i}-1\right).
\end{eqnarray}
Note that this expression is consistent with the result (\ref{energy3}) for $E(t)$
(taking into account that $G(0)=0$ and $S(0)=1$).
If the new frequency $\omega$ corresponds to an ergodic configuration ($\omega\le\omega_c$) then  the excess energy $\Delta E$
is eventually dissipated into the bath, and the oscillator  (now with the new frequency $\omega$)
returns to thermal equilibrium with the average energy $E=k_BT$. 
On the other hand, if the new frequency corresponds to a nonergodic configuration ($\omega> \omega_c$)
then  only a part of the excess energy $\Delta E$ dissipates into the bath. The dissipated energy
$E_{diss}=E(0^+)-E_{cs}(t)$ oscillates in time. Being averaged over time,  it takes the form
\begin{eqnarray}
\overline E_{diss}&=&E(0^+)-\overline E_{cs}=f(\lambda,\mu)\,\Delta E
\label{E_diss}
\end{eqnarray}
with 
\begin{eqnarray}
f(\lambda,\mu)=1-
\frac{1}{2} \left[
(\lambda/\beta)^2
+\lambda
\right]\,
(\omega_0G_0)^2,
\end{eqnarray}
and $\lambda>\lambda_c$.
According to Eq. (\ref{E_diss}), the function $f(\lambda,\mu)$ has the meaning of the fraction of the initial
excess energy $\Delta E$ eventually dissipated into the 
bath, 
so one  expects $0<f(\lambda,\mu)\le 1$.

Consider specifically the case $\mu=1$ when the expressions for $\beta$ and $G_0$, see Sec. XI, are less bulky,
\begin{eqnarray}
\omega_0\,G_0=\frac{8\,\lambda-4}{(4\,\lambda-1)^{3/2}},\quad
\beta=\frac{2\lambda}{\sqrt{4\lambda-1}},
\label{mu1}
\end{eqnarray}
and $\lambda_c=1-\mu/2=1/2$.
For that case $f(\lambda)$ takes the form
\begin{eqnarray}
f(\lambda)=\frac{24\lambda^2-12 \lambda+1}{(4 \lambda-1)^3}.
\end{eqnarray}
One observes that $f(\lambda)$ takes the value $1$ for $\lambda=\lambda_c$ and monotonically  decreases as $1/\lambda$.
This means that for a nonergodic configuration with $\lambda>\lambda_c$ only a part $f\,\Delta E$ of the excess energy
$\Delta E$ is dissipated into the bath at long times, another part $(1-f)\Delta E$ remains localized in the oscillator.
The localized energy monotonically increases with $\lambda$ and can be arbitrary large.

In our view, the ability to keep permanently the energy exceeding the thermal equilibrium value $k_BT$, in  violation of
the equipartition theorem, is the key property
of the nonergodic oscillator. It is tempting to view configurations with $\lambda\le \lambda_c$ and $\lambda>\lambda_c$
as two ``phases" and to consider 
ergodic to nonergodic transitions 
$\lambda_i\to\lambda$ (with $\lambda_i<\lambda_c$) as a phase transition with  the dimensionless order parameter 
\begin{eqnarray}
\eta(\lambda)=\frac{\overline E_{cs}(\lambda)-k_BT}{k_BT}.
\end{eqnarray}
In the ergodic phase ($\lambda\le\lambda_c$) the order parameter vanishes, while in the nonergodic phase
($\lambda>\lambda_c$) it is non-zero and increases with $\lambda$, 
\begin{eqnarray}
\eta(\lambda)=
\begin{cases}
0, & \text{if $\lambda\le\lambda_c$},\\
\frac{1}{4}\left(
\frac{\lambda}{\lambda_i}-1\right)
\left[
\left(\frac{\lambda}{\beta}
\right)^2
+
\lambda
\right]\,(\omega_0 G_0)^2, & \text{if $\lambda>\lambda_c$}.
\end{cases}
\end{eqnarray}
For $\mu=1$, when $G_0$ and $\beta$ are given by Eq. (\ref{mu1}) (and $\lambda_c=1/2$), the explicit dependence
of the order parameter on $\lambda$ is
\begin{eqnarray}
\eta(\lambda)=
\begin{cases}
0, & \text{if $\lambda\le \lambda_c$},\\
c(\lambda,\lambda_i)\,(\lambda-\lambda_c)^2, & \text{if $\lambda>\lambda_c$},
\end{cases}
\label{FT}
\end{eqnarray}
with
\begin{eqnarray}
c(\lambda,\lambda_i)=
\frac{4(8\lambda-1)}{(4\lambda-1)^3}\,
\left(\frac{\lambda}{\lambda_i}-1
\right).
\end{eqnarray}
Since  $\eta (\lambda)$ and its  first derivatives $\eta'(\lambda)$ are continuous and  the second derivative $\eta''(\lambda)$ is discontinuous at 
 $\lambda=\lambda_c$, see Fig. 9, there is  a resemblance between  
ergodic to nonergodic transitions and conventional phase transitions of second order. 
Note that the presented study is limited to the case when the switching $\lambda_i\to\lambda$ occurs
instantaneously, and there is no reason to believe that the exponent $2$ in Eq. (\ref{FT})  would be the same if the switching takes a finite time.

\begin{figure}[t]
\includegraphics[height=5.5cm]{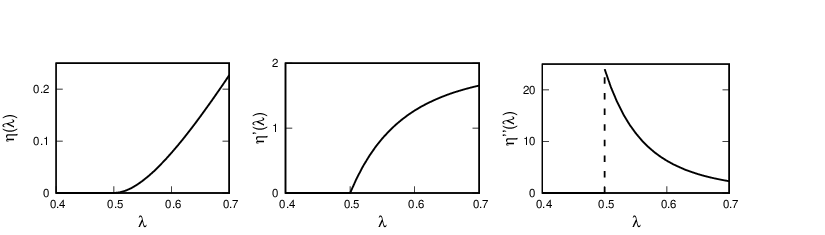}
\caption{ The order parameter $\eta(\lambda)$, its first  $\eta'(\lambda)$ and
second $\eta''(\lambda)$ derivatives for $\mu=1$ and $\lambda_i=0.25$ near the critical point $\lambda_c=0.5$.
}
\end{figure}

\section{Conclusion}
In this paper  we have evaluated the relaxation and correlation functions for a Brownian  oscillator 
described by the generalized Langevin equation with the dissipation kernel of the form (\ref{K}). 
The oscillator may have both ergodic and nonergodic configurations (for $\mu<2$) , or only nonergodic configurations (for $\mu\ge 2$).
In ergodic configurations, which correspond to lower oscillator frequencies $\omega\le \omega_c=\sqrt{1-\mu/2}\,\omega_0$,
the oscillator relaxes to thermal equilibrium with the external bath. In nonergodic configurations, corresponding
to higher oscillator frequencies $\omega>\omega_c$, the oscillator does not reach thermal equilibrium
(unless prepared in equilibrium initially), but evolves to non-equilibrium cyclostationary states
in which the average oscillator's energy oscillates with time and exceeds the equilibrium value $k_BT$ prescribed by the equipartition theorem.

In general, we observed that nonergodic configurations
emerge when the spectrum of the bath's modes is bounded from the above. That is not the case for gas-like environments,
but characteristic for lattices. The specific model considered here corresponds to an isotope atom embedded
in an infinite or semi-infinite harmonic chain (Rubin's model) and subjected to the external harmonic potential.

In the limit of zero oscillator frequency $\omega\to 0$, or $\lambda=(\omega/\omega_0)^2\to 0$,
the presented results  recover that for Rubin's model.
In particular, for the relaxation function $R(t)$, which is also the normalized 
velocity autocorrelation 
function in equilibrium, 
in the limit
$\lambda\to 0$ the presented results
take the form
\begin{eqnarray}
R(t)&=&
\begin{cases}
    R_e(t), & \text{if $\mu<  2$},\\
R_e(t)+R_0\,\cos(\omega_* t), & \text{ if $\mu\ge 2$}.
  \end{cases}
  \label{Rubin1}
 \end{eqnarray}
This reflects that for $\mu<2$ the condition of nonergodic relaxation $\lambda> \lambda_c=1-\mu/2$
cannot by satisfied when $\lambda\to 0$, and that for $\mu\ge 2$ the relaxation is nonergodic for any $\lambda$,
 including the limit $\lambda\to 0$. As follows from Eq.(\ref{R2}), the ergodic component of $R(t)$ in the limit $\lambda\to 0$ reads
\begin{eqnarray}
R_e(t)=\frac{4\mu}{\pi}\,\int_{0}^{1}
\frac{\sqrt{1-x^2}\,\cos (x\,\omega_0 t)\,dx}{4(1-\mu)\,x^2
+\mu^2}.
\label{Rubin2}
\end{eqnarray}
In particular, for $\mu=2$, which corresponds to the case of a tagged atom in a bulk of the uniform harmonic chain,  we recover the well-known result~\cite{Zwanzig}
\begin{eqnarray}
R(t)=R_e(t)=\frac{2}{\pi}\,\int_{0}^{1}
\frac{\cos (x\,\omega_0 t)\,dx}{\sqrt{1-x^2}}=J_0(\omega_0 t).
\nonumber
\end{eqnarray}
For the amplitude and frequency 
of the nonergodic component we get
from Eqs. (\ref{poles5555}) and 
(\ref{R2}) in the limit $\lambda\to 0$

\begin{eqnarray}
R_0=\frac{\mu-2}{\mu-1},\quad
\omega_*=\frac{\mu}{2\sqrt{\mu-1}}\,\omega_0.
\label{Rubin3}
\end{eqnarray}
As expected, Eqs. (\ref{Rubin1})-(\ref{Rubin3}) coincide with the known results for Rubin's model, see Eq. (A28)
in Ref.~\cite{Rubin} (note that parameter $Q$  of Ref.~\cite{Rubin} and parameter $\mu$ used in this paper are related as $Q=2/\mu-1$).

The relaxation functions evaluated in the paper allow us to describe the evolution of the oscillator for $t>0$
with a frequency which is either fixed or switched instantaneously  
at $t=0$. The latter setting allows one to study a particular type of ergodic to nonergodic transitions.
An interesting extension would be to study such transitions when the oscillator frequency is tuned continuously during a finite time.  
Such an extension would be  of interest, in particular,  from the perspective  of the fluctuation theorems~\cite{Seifert_review}. 
Their proofs often assume that the system is ergodic for all values of the tunable parameter $\lambda(t)$~\cite{J,Seifert}.
Not much is currently known about what happen if that is not the case (see, however,~\cite{Hasegawa}).
Another interesting application is Brownian engines~\cite{BE,BE2,BE3}.
A nonergodic oscillator in a cyclostationary state may store an arbitrary amount of energy.
That property may be used beneficially in designing Brownian machines.
More generally,
the parametric erodic to nonergodic transitions may be of interest because
algorithms involving  periodically correlated  (cyclostationary) processes in nonergodic configurations are often
advantageous relative to those based on stationary  processes characteristic for ergodic regimes~\cite{Serpedin}.

It might be tempting to seek implications of the presented results in the context of experiments with colloidal particles held in optical traps.
In a version called the capture experiment the strength of the optical trap  (the oscillator frequency in our model)
is changed instantaneously, and the relaxation of the particle's position and velocity is recorded~\cite{traps1,traps2}.
This is precisely the setting described by the relaxation functions obtained in this paper. However, when the bath
is formed by a lattice, nonergodic configurations correspond to frequencies of order of $\omega_0$ which,
even for soft lattices,  is several orders of magnitude higher than frequencies used in optical
trap experiments with colloidal particles in gaseous and aqueous environments.

\renewcommand{\theequation}{A\arabic{equation}}
  \setcounter{equation}{0}  

\section*{Appendix A: Evaluation of $I_0(t)$}

In this Appendix we evaluate the integral $I_0$ defined by Eq. (\ref{I0_def}).
Consider the integral
\begin{eqnarray}
I_0^+(t)=\frac{1}{2\pi i}\,\int_{\Gamma_0^+} e^{st} \, \tilde G(s)\,ds=
\frac{1}{\pi i}\,\int_{\Gamma_0^+}
\frac{e^{st}\,ds}{(2-\mu)\,s^2+\mu\,s\,f_1(s)+2\,\lambda\,\omega_0^2}
\end{eqnarray} 
along the right shore of the branch cut, see Fig. 1, in the direction from $i\omega_0$ to $-i\omega_0$. Recall that $f_1(t)$
is the physical branch of the function $f(s)=\sqrt{s^2+\omega_0^2}$ and can be evaluated as 
\begin{eqnarray}
f_1(s)=\sqrt{r_1r_2}\, e^{i\frac{\theta_1+\theta_2}{2}}
\end{eqnarray}
in terms of polar coordinates defined in Fig.1 with both  polar angles $\theta_{1,2}$ in the range
$(-3\pi/2,\pi/2]$, see 
Eqs. (\ref{f}) and (\ref{f1}).
For points $s\in \Gamma_0^+$ one can use the parametrization $s=iy+\epsilon$ with $-\omega_0\le y\le\omega_0$;
then in the limit $\epsilon\to 0$ one finds:
\begin{eqnarray}
\theta_1=-\frac{\pi}{2},\quad
\theta_2=\frac{\pi}{2},\quad
r_1=\omega_0-y,\quad
r_2=\omega_0+y,
\end{eqnarray}
Therefore
\begin{eqnarray}
f_1(s)=\sqrt{\omega_0^2-y^2}\quad
\mbox{for}\quad
s\in \Gamma_0^+,
\end{eqnarray}
and 
\begin{eqnarray}
I_0^+(t)=
-\frac{1}{\pi}\,\int_{-\omega_0}^{\omega_0}
\frac{e^{i yt}\,dy}{(\mu-2)\,y^2+i\mu\,y\,\sqrt{\omega_0^2-y^2}+2\,\lambda\,\omega_0^2}.
\end{eqnarray} 
In a similar manner we can evaluate the integral
\begin{eqnarray}
I_0^-(t)=\frac{1}{2\pi i}\,\int_{\Gamma_0^-} e^{st} \, \tilde G(s)\,ds=
\frac{1}{\pi i}\,\int_{\Gamma_0^-}
\frac{e^{st}\,ds}{(2-\mu)\,s^2+\mu\,s\,f_1(s)+2\,\lambda\,\omega_0^2}
\end{eqnarray} 
along the left shore of the branch cut in the direction from $-i\omega_0$ to $i\omega_0$.
In that case
\begin{eqnarray}
\theta_1=-\frac{\pi}{2},\quad
\theta_2=-\frac{3\pi}{2},\quad
r_1=\omega_0-y,\quad
r_2=\omega_0+y.
\end{eqnarray}
This gives 
\begin{eqnarray}
f_1(s)=-\sqrt{\omega_0^2-y^2}\quad
\mbox{for}\quad
s\in \Gamma_0^-,
\end{eqnarray}
and 
\begin{eqnarray}
I_0^-(t)=
\frac{1}{\pi}\,\int_{-\omega_0}^{\omega_0}
\frac{e^{i yt}\,dy}{(\mu-2)\,y^2-i\mu\,y\,\sqrt{\omega_0^2-y^2}+2\,\lambda\,\omega_0^2}.
\end{eqnarray} 
Considering the sum
$I_0=I_0^++I_0^-$ and evaluating its real and imaginary parts, one finds  that the latter is zero  due to symmetry, and the result is
\begin{eqnarray}
I_0(t)=-\frac{4\mu}{\pi}\,\int_{0}^{\omega_0}
\frac{y\sqrt{\omega_0^2-y^2}\,\sin (y t)\,dy}{4(1-\mu)\,y^4
+[4\lambda (\mu-2)+\mu^2]\,\omega_0^2\, y^2+4\lambda^2\omega_0^4}.
\end{eqnarray}
It is convenient to present the result using the dimensionless integration variable $x=y/\omega_0$,
\begin{eqnarray}
I_0(t)=-\frac{4\mu}{\pi\omega_0}\,\int_{0}^{1}
\frac{x\,\sqrt{1-x^2}\,\sin (x\,\omega_0 t)\,dx}{4(1-\mu)\,x^4
+[4\lambda(\mu-2)+\mu^2]\, x^2+4\lambda^2}.
\label{I0_result}
\end{eqnarray}
This result holds for arbitrary $\mu$.

\renewcommand{\theequation}{B\arabic{equation}}
  \setcounter{equation}{0}  

\section*{Appendix B: Evaluation of residues}

Here we evaluate the residues in Eqs. (\ref{baux00}) and (\ref{daux00}) for the relaxation function $G(t)$.
One can verify that for arbitrary $\mu$ the poles $s_{1,2}=\pm i\omega_*$ of the transform $\tilde G(s)$
are of order one  (simple poles). Then  the residue of $e^{st}\tilde G(s)$ at $s_1$ is evaluated as follows:
\begin{eqnarray}
Res[e^{st} \tilde G(s), s_1]=\lim_{s\to s_1} e^{st}\tilde G(s)\,(s-s_1)=e^{i\omega_* t}\lim_{s\to i\omega_*} \tilde G(s) (s-i\omega_*)=
e^{i\omega_* t}\lim_{s\to i\omega_*}\frac{2 (s-i\omega_*)}{(2-\mu)\,s^2+\mu\,s f_1(s)+2\lambda\omega_0^2}.\nonumber
\end{eqnarray}
Applying L`Hospital's Rule  yields
\begin{eqnarray}
Res[e^{st} \tilde G(s), s_1]=
e^{i\omega_* t}\lim_{s\to i\omega_*}\frac{2 f_1(s)}{\mu\,s^2+2(2-\mu)s\,f_1(s)+\mu\,f_1^2(s)}.
\end{eqnarray}
Then, recalling Eq. (\ref{aux1}), $f_1(i \omega_*)= i\sqrt{\omega_*^2-\omega_0^2}$,
one gets
\begin{eqnarray}
Res[e^{st} \tilde G(s), s_1]=\frac{2i\,\sqrt{\omega_*^2-
\omega_0^2}}{\mu(\omega_0^2-2\omega_*^2)-2(2-\mu)\omega_*\,\sqrt{\omega_*^2-\omega_0^2}}\,e^{i\omega_* t}
\end{eqnarray}
Similarly, for the residue at the second pole $s_2=-i\omega_*$ we obtain
\begin{eqnarray}
Res[e^{st} \tilde G(s), s_2]
=\frac{-2i\,\sqrt{\omega_*^2-
\omega_0^2}}{\mu(\omega_0^2-2\omega_*^2)-2(2-\mu)\omega_*\,\sqrt{\omega_*^2-\omega_0^2}}\, e^{-i\omega_*t}.
\end{eqnarray}
The sum of residues is
\begin{eqnarray}
Res[e^{st} \tilde G(s), s_1]+
Res[e^{st} \tilde G(s), s_2]
=\frac{4\,\sqrt{\omega_*^2-
\omega_0^2}}{\mu\,(2\omega_*^2-\omega_0^2)+2(2-\mu)\omega_*\,\sqrt{\omega_*^2-\omega_0^2}}\,\sin(\omega_* t),
\end{eqnarray}
These expressions hold for arbitrary $\mu$, although the frequency $\omega_*=\omega_*(\lambda)$ has different forms for $\mu=1$ and for $\mu\ne 1$.
Introducing the dimensionless function
$\beta(\lambda)=\omega_*(\lambda)/\omega_0$, 
the above expression can be written as
\begin{eqnarray}
Res[e^{st} \tilde G(s), s_1]+
Res[e^{st} \tilde G(s), s_2]
=\frac{1}{\omega_0}\,\frac{4\,\sqrt{\beta^2-1}}{\mu\,(2\beta^2-1)+2(2-\mu)\beta\,\sqrt{\beta^2-1}}\,\sin(\omega_* t).
\end{eqnarray}


\end{document}